%% file: main.tex
\documentclass[journal=jpcafh,manuscript=article]{achemso}

\usepackage{chemformula} 
\usepackage[T1]{fontenc} 
\usepackage{epstopdf}
\usepackage{hyperref}
\usepackage{lineno,hyperref,amsmath}
\usepackage{xcolor}

\usepackage{xr}
\makeatletter
\newcommand*{\addFileDependency}[1]{
  \typeout{(#1)}
  \@addtofilelist{#1}
  \IfFileExists{#1}{}{\typeout{No file #1.}}
}
\makeatother

\newcommand*{\myexternaldocument}[1]{%
    \externaldocument{#1}%
    \addFileDependency{#1.tex}%
    \addFileDependency{#1.aux}%
}

\myexternaldocument{supp_information}

\author{Sergio A. Urz\'ua}
\affiliation[USM]
{Department of Mechanical Engineering, Universidad T\'ecnica Federico Santa Mar\'ia, Valpara\'iso, Chile}
\author{Perla Y. Sauceda-Olo\~no}
\affiliation[Clem]
{Department of Chemistry, Clemson University, Clemson, SC}
\author{Carlos D. Garc\'ia}
\affiliation[Clem]
{Department of Chemistry, Clemson University, Clemson, SC}
\author{Christopher D. Cooper}
\affiliation[USM]
{Department of Mechanical Engineering, Universidad T\'ecnica Federico Santa Mar\'ia, Valpara\'iso, Chile}
\alsoaffiliation[CCTVal]
{Centro Cient\'ifico Tecnol\'ogico de Valpara\'iso, Valpara\'iso, Chile}
\email{christopher.cooper@usm.cl}

\title[]
  {Predicting the orientation of adsorbed proteins steered with electric fields using a simple electrostatic model 
}

\abbreviations{IR,NMR,UV}

\begin{document}

\begin{tocentry}
\includegraphics{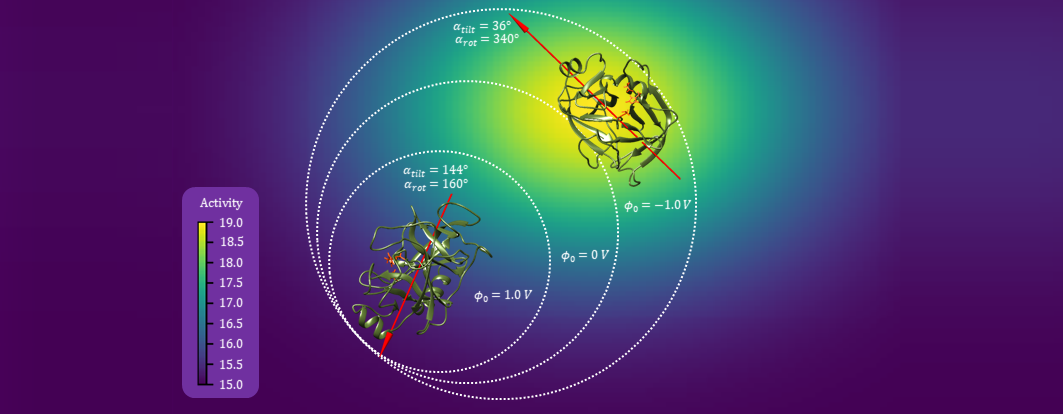}
\end{tocentry}
\newpage
\section{Abstract}
\input{abstract}

\newpage
\section{Introduction}
\input{intro}

\section{Materials and Methods}
\input{methods}

\section{Results}
\input{results}

\section{Discussion}
\input{discussion}

\section{Conclusions}
\input{conclusions}

\section{Acknowledgements}
Financial support for this project has been provided in part by Universidad T\'ecnica Federico Santa Mar\'ia (project PI-LIR-2020-10) and Clemson University. CDC acknowledges the support from CCTVal through ANID PIA/APOYO AFB180002 and computational resources.

\section{Conflicts of Interest}
Authors declare no conflict of interest related to the material.





\bibliography{main}

\end{document}


\maketitle	

\section{Verification of the numerical model with an analytical solution}

\subsection{A closed expression for spheres with Legendre polynomials}

\begin{figure}[!h]
    \centering
    \includegraphics[scale = 0.51]{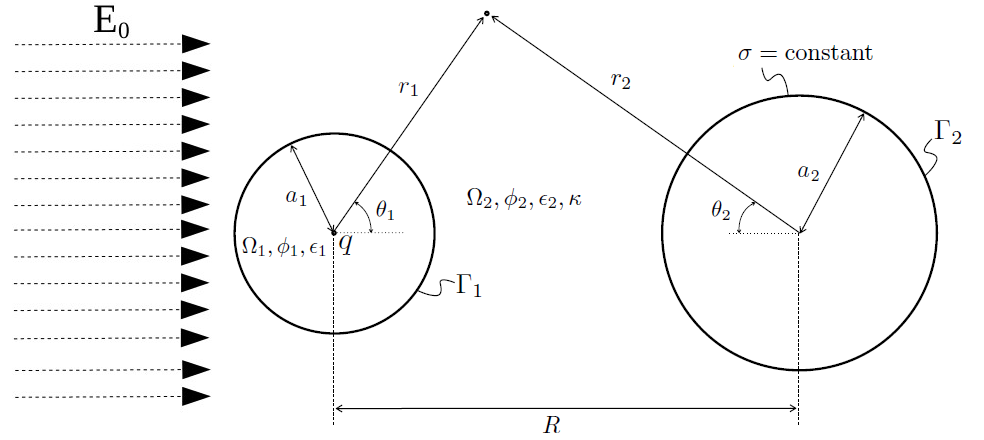}
    \caption{Sketch of a spherical molecule interacting with a charged sphere under an external electric field.}
    \label{fig:spheres_analytical}
\end{figure}

This section describes a derivation towards a closed expression for the electrostatic potential of a spherical molecule with a centered charge interacting with a charged sphere, under a constant electric field (see Fig. \ref{fig:spheres_analytical}). This expression is based on Legendre polynomials. We realize this is not a physically realistic setup, as it does not consider ionic screening of the external field, however, it is useful to verify the numerical implementation, as shown in Fig \ref{fig:sphere_convergence}. 

Considering azimuthal symmetry, the solution of the coupled system in Eq. (1) can be written as \cite{carnie1994electrical}

\begin{eqnarray}\label{eq:an_1}
        \phi_{1} = \sum_{j = 0}^{\infty} C_{n}r^{n}P_{n}(\cos{\theta_{1}}) + \dfrac{q_{k}}{4\pi\epsilon_{1}}\sum_{n = 0}^{\infty}\dfrac{r^{n}_{k}}{r^{n + 1}}P_{j}(\cos{\theta_{1}})\nonumber\\
        \phi_{2} = \sum_{n = 0}^{\infty} a_{n}K_{n}(\kappa r_{1})P_{n}(\cos{\theta_{1}}) + \sum_{n = 0}^{\infty} b_{n}K_{n}(\kappa r_{2})P_{n}(\cos{\theta_{2}}) \nonumber \\
        \phi_{e} = -E_{0}r_{1}P_{1}(\cos{\theta_{1}})\nonumber\\
        \phi_{1} = \phi_{2} + \phi_{e} \hspace{1cm}; \hspace{0.2cm} \Gamma(r = a_{1}) \nonumber\\
        \epsilon_{1} \dfrac{\partial \phi_{1}}{\partial n} = \epsilon_{2}\left(\dfrac{\partial \phi_{2}}{\partial n} + \dfrac{\partial \phi_{e}}{\partial n}\right) \hspace{0.75cm} \nonumber\\
         \hspace{0.8cm} -\epsilon_{2}\left(\dfrac{\partial \phi_{2}}{\partial n} + \dfrac{\partial \phi_{e}}{\partial n}\right) = \sigma_{0} \hspace{0.2cm} ; \hspace{0.2cm} \Gamma_{2}(r = a_{2})
\end{eqnarray}

where $P_{n}$ and $K_{n}$ are the $n^{th}$ Legendre polynomial and modified spherical Bessel function of second kind, respectively. Using the addition theorem\cite{marvcelja1977role}, $\phi_2$ and $\phi_e$ are written as
\begin{eqnarray}
     \phi_{2} = \sum_{n = 0}^{\infty} a_{n}K_{n}(\kappa r_{1})P_{n}(\cos{\theta_{1}})+
     \sum_{n = 0}^{\infty} b_{n}\sum_{m = 0}^{\infty}(2m + 1)B_{nm}i_{m}(\kappa r_{1})P_{m}(\cos{\theta_{1}}) \nonumber \\
      \phi_{2} = \sum_{n = 0}^{\infty} b_{n}K_{n}(\kappa r_{2})P_{n}(\cos{\theta_{2}}) + 
      \sum_{n = 0}^{\infty} a_{n} \sum_{m = 0}^{\infty}(2m + 1)B_{nm}i_{m}(\kappa r_{1})P_{m}(\cos{\theta_{1}}) \nonumber \\
      \phi_{e} = -E_{0}r_{1}P_{1}(\cos{\theta_{1}}) = E_{0}r_{2}P_{1}(\cos{\theta_{2}})-E_{0}R\nonumber\\
\end{eqnarray}
where $i_{m}$ corresponds to the modified spherical Bessel function of the first kind. Moreover, $B_{nm}$ is defined as
\begin{equation}
    B_{nm} = \sum_{\nu = 0}^{\infty}A_{nm}^{\nu}K_{n + m - 2\nu}(\kappa R)
\end{equation}
where $R$ is the center-to-center distance, and $A_{nm}^{\nu}$ is related to the Gamma function, which is given by the following expression.

\begin{equation}
    A_{nm}^{\nu}=\dfrac{\Gamma_{n-\nu+0.5}\Gamma_{m-\nu+0.5}\Gamma_{\nu+0.5}(n+m-nu)!(n+m-2\nu+0.5)}{\pi \Gamma_{m+n-\nu+1.5}(n-\nu)!(m-\nu)!\nu!}
\end{equation}

Evaluating the expressions in Eq. \eqref{eq:an_1} at $\Gamma_1$, and applying the interface conditions on the potential and electric displacement yields
\begin{align} \label{eq:an_2}
  C_{j}a_{1}^{j} + \dfrac{q_{k}}{4\pi\epsilon_{1}}\left(\dfrac{r_{k}^{j}}{a_{1}^{j+1}}\right) &= a_{j}K_{j}(\kappa a_{1}) + \sum_{n = 0}^{\infty} b_{n}(2j + 1)B_{nj}i_{j}(\kappa a_{1}) - E_{0}r_{1}\delta_{1j} \nonumber\\ 
  \dfrac{\epsilon_{1}}{\epsilon_{2}}\left(C_{j}ja_{1}^{j-1} - \dfrac{q_{k}(j+1)}{4\pi\epsilon_{1}}\left(r_{k}^{j}a_{1}^{-(j+2)}\right)\right) =& a_{j}\kappa K'_{j}(\kappa a_{1}) + 
           + \sum_{n = 0}^{\infty} b_{n}(2j + 1)B_{nj}\kappa i'_{j}(\kappa a_{1}) - E_{0}\delta_{1j}.
\end{align}
Rearranging terms, this leads to
\begin{align}\label{eq:an_3}
    &C_{j} = \dfrac{1}{a_{1}^{j}}\left(a_{j}K_{j}(\kappa a_{a}) + \sum_{n=0}^{\infty}b_{n}(2j+1)B_{nj}i_{j}(\kappa a_{1}) - E_{0}\delta_{1j} - \dfrac{q_{k}}{4\pi\epsilon_{1}}\dfrac{r_{k}^{j}}{a_{1}^{j+1}}\right) \nonumber\\
    &\sum_{n = 0}^{\infty}\left(a_{n}\left[\kappa K'(\kappa a_{1}) - \dfrac{\epsilon_{1}}{\epsilon_{2}}\dfrac{n}{a_{1}}K_{n}(\kappa a_{1})\right]\delta_{nj} + b_{n}(2j+1)B_{nj}\left[\kappa i_{j}'(\kappa a_{1}) - \dfrac{\epsilon_{1}}{\epsilon_{2}}\dfrac{j}{a_{1}}i_{j}(\kappa a_{1})\right]\right) \nonumber\\ 
    &= - \dfrac{q_{k}(2j+1)}{4\pi\epsilon_{2}}\left(\dfrac{r_{k}^{j}}{a_{1}^{j+2}}\right) - E_{0}\left(\dfrac{\epsilon_{1}}{\epsilon_{2}}j - 1\right)\delta_{1j}
\end{align}

Now, applying the conditions on $\Gamma_2$ gives
\begin{equation}\label{eq:an_4}
    \sum_{n=0}^{\infty}b_{n}\kappa K'_{n}(\kappa a_{2})\delta_{nj} + a_{n}(2j+1)B_{nj}\kappa i'_{j}(\kappa a_{2}) = -\dfrac{\sigma_{0}}{\epsilon_{2}}\delta_{0j}-E_{0}\delta_{1j}
\end{equation}
We can now use Eqs. \eqref{eq:an_1}, \eqref{eq:an_2}, and \eqref{eq:an_3} to generate a linear system for every index $j$, as
\begin{eqnarray}\label{eq:an_matrix}
\begin{bmatrix}
  I_{jn} & L_{jn} \\
  M_{jn} & I_{jn} \\
 \end{bmatrix}
 \begin{bmatrix}
  A_{n} \\
  B_{n}  \\
 \end{bmatrix}
    =  
  \begin{bmatrix}
   -E_{0}\left(\dfrac{\epsilon_{1}}{\epsilon_{2}}j - 1\right)\delta_{1j} - \dfrac{q_{k}(2j + 1)}{4\pi\epsilon_{2}}\dfrac{r_{k}^{j}}{a_{1}^{j + 2}}\\
  -E_{0}\delta_{1j} -\dfrac{\sigma_{0}}{\epsilon_{2}}\delta_{0j} \\
 \end{bmatrix}
\end{eqnarray}
where $A_{n}$, $B_{n}$, $I_{jn}$, $L_{jn}$ y $M_{jn}$ are:
\begin{eqnarray}
    A_{n} = a_{n}\left(\kappa K'_{n}(\kappa a_{1}) - \dfrac{\epsilon_{1}}{\epsilon_{2}}\dfrac{n}{a_{1}}K_{n}(\kappa a_{1})\right)\nonumber\\
    B_{n} = b_{n}\kappa K'_{n}(\kappa a_{2})\nonumber\\
    I_{jn} = \delta_{jn} \nonumber\\
    M_{jn} = \dfrac{(2j + 1)B_{nj}\kappa i'_{j}(\kappa a_{2})}{\left(\kappa K'_{n}(\kappa a_{1}) - \dfrac{\epsilon_{1}}{\epsilon_{2}}\dfrac{n}{a_{1}}K_{n}(\kappa a_{1})\right)}\nonumber\\
    L_{jn} = (2j + 1)B_{nj}\left(\dfrac{i'_{j}(\kappa a_{1})}{K'_{n}(\kappa a_{2})} - \dfrac{\epsilon_{1}}{\epsilon_{2}}\dfrac{j}{a_{1}}\dfrac{i_{j}(\kappa a_{1})}{\kappa K'_{n}(\kappa a_{2})}\right)
\end{eqnarray}
Having $A_n$ and $B_n$, we can use Eq. \eqref{eq:an_3} to compute $C_j$, and then Eq. \eqref{eq:an_1} to obtain the electrostatic potential anywhere in the domain.

\subsection{Numerical verification}

\begin{figure} 
    \centering
    \includegraphics[width=0.45\textwidth]{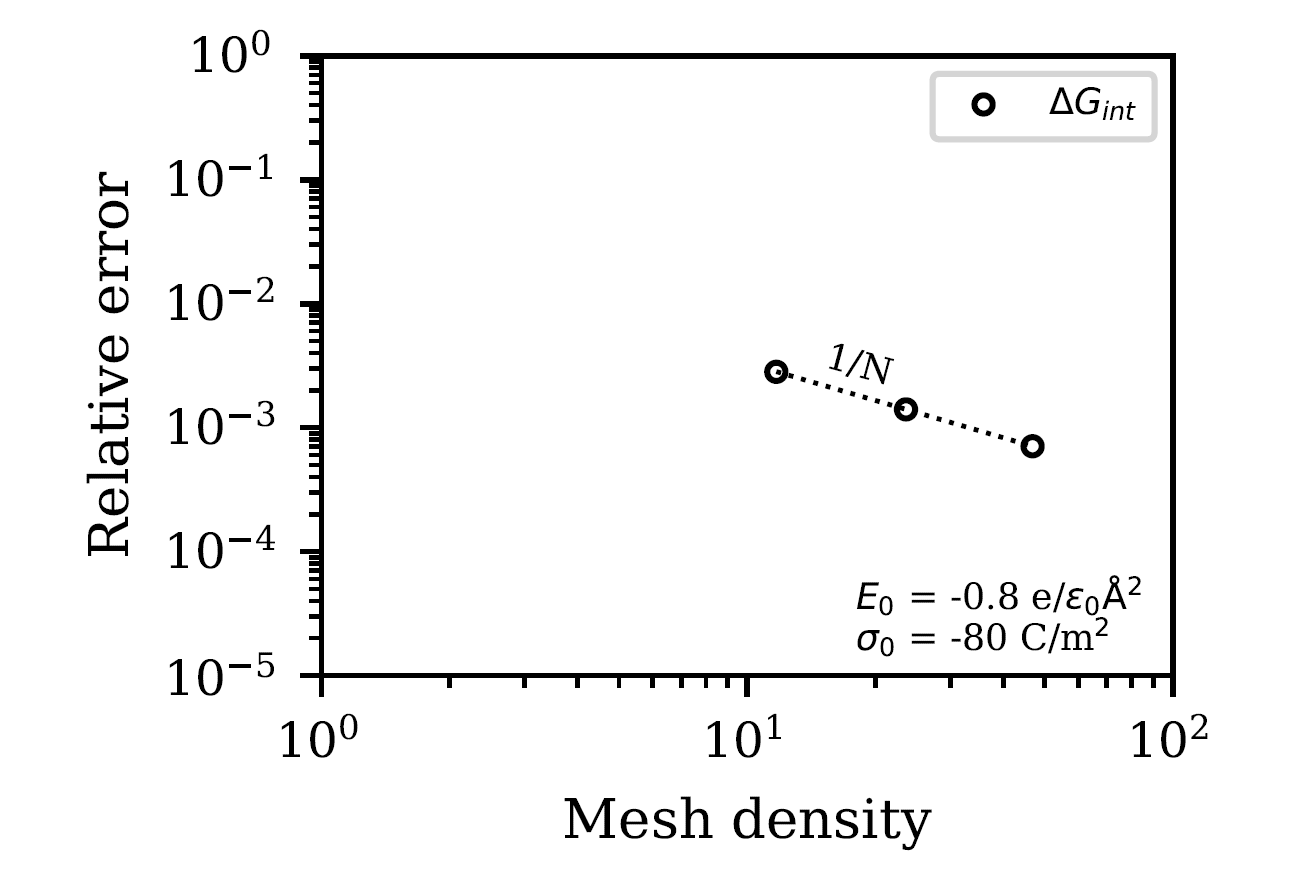}
    \caption{Mesh convergence for a spherical molecule interacting with a charged sphere.}
    \label{fig:sphere_convergence}
\end{figure}

Here, we use the derivation that led to Eq. \eqref{eq:an_matrix} and the setup in Fig. \ref{fig:spheres_analytical} to validate our numerical implementation on PyGBe, with $\sigma_0=-80$ C/m$^2$, $E_0=-0.8$ $e^-/\epsilon_0$\AA$^2$, $a_{1} = a_{2} = 4$ \AA, and $R = 14$ \AA. Table \ref{table:verification_param} shows the numerical parameters of these runs. In this case, the analytical solution computed with Eq.~\eqref{eq:an_matrix} was $\Delta G_{solv} = -652.43$ kcal/mol and $G_{surf} = -2.88\cdot 10^{8}$ kcal/mol. Figure \ref{fig:sphere_convergence} plots the error convergence with mesh density, where we see the expected 1/$N$ behavior.~\cite{CooperBardhanBarba2014}

\begin{table} [!h]
\centering
\caption{PyGBe parameters for verification runs.}
\label{table:verification_param}
\begin{tabular}{ccccccc}
\hline
\multicolumn{3}{c}{\# Gauss points} & \multicolumn{3}{c}{Treecode} & GMRES     \\ \hline
$K$    & $K_{fine}$    & $N_{k}$    & $N_{crit}$  & P   & $\theta$ & tol.      \\ \hline
4      & 37            & 9          & 300         & 15  & 0.5      & $10^{-9}$ \\ \hline
\end{tabular}
\end{table}

\section{Simulations for trypsin adsorbed on a charged surface.}

\subsection{Numerical parameter selection}

The interaction energy in Eq.(7) of the main text is a difference between large numbers. Then, we need to be extremely careful about the accuracy of our simulations. We performed simulations of trypsin placed 2~\AA~away from a 100$\times$100$\times$10 \AA~ surface, uncharged ($\sigma_0$=0 C/m$^2$), under a field of $\phi_0=\pm1.0$ V and $\phi_0=0$ V, and studied the mesh convergence. In this case, the numerical parameters of PyGBe were specially tight (see Table \ref{table:convergence_trypsin_param}) to ensure that the numerical approximations of the integration and treecode did not contaminate the mesh convergence. We parameterized the charge and radii of trypsin with \texttt{pdb2pqr}~\cite{dolinsky2004pdb2pqr} considering pH=8, and triangulated its surface with \texttt{Nanoshaper}.~\cite{decherchi2013general} 

\begin{figure}
    \centering
    \includegraphics[width=0.45\textwidth]{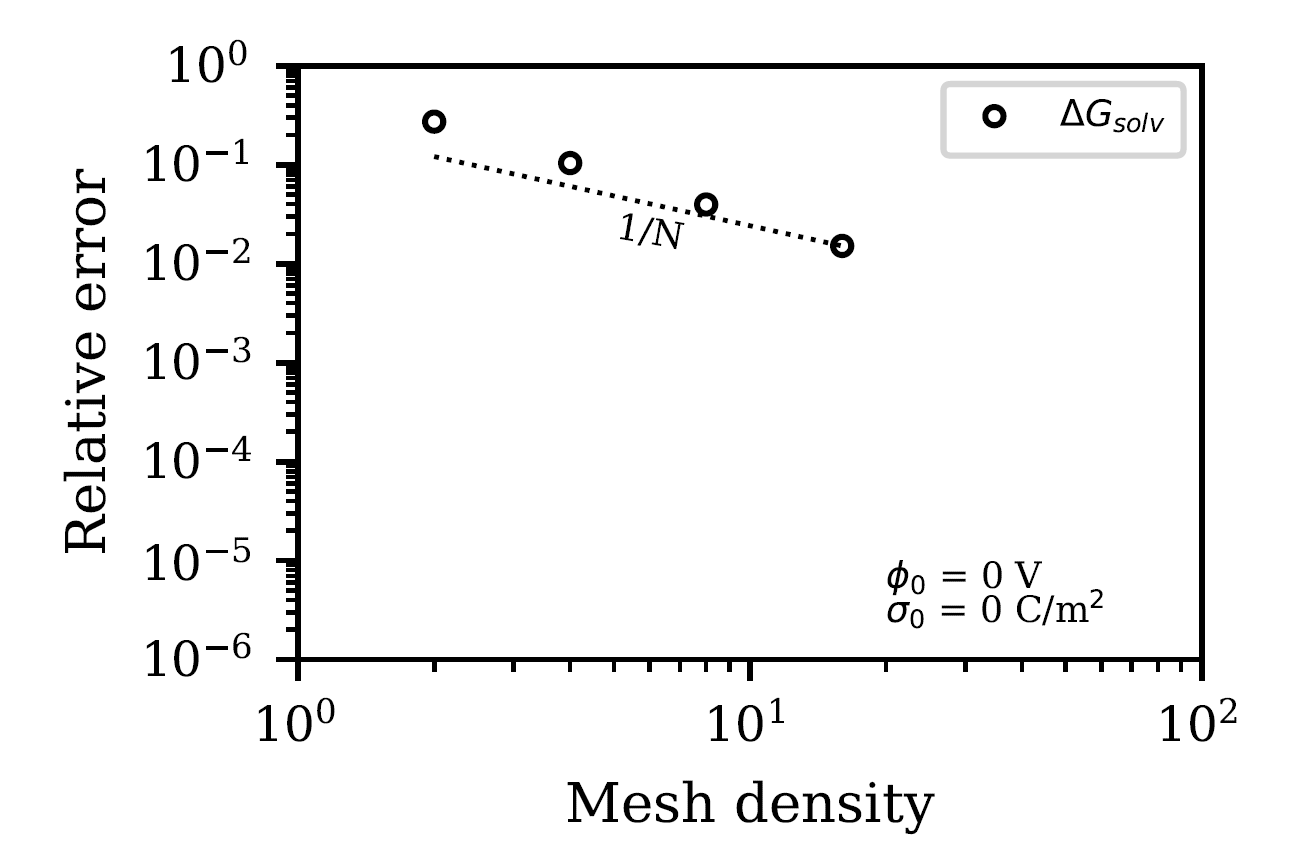}
    \includegraphics[width=0.45\textwidth]{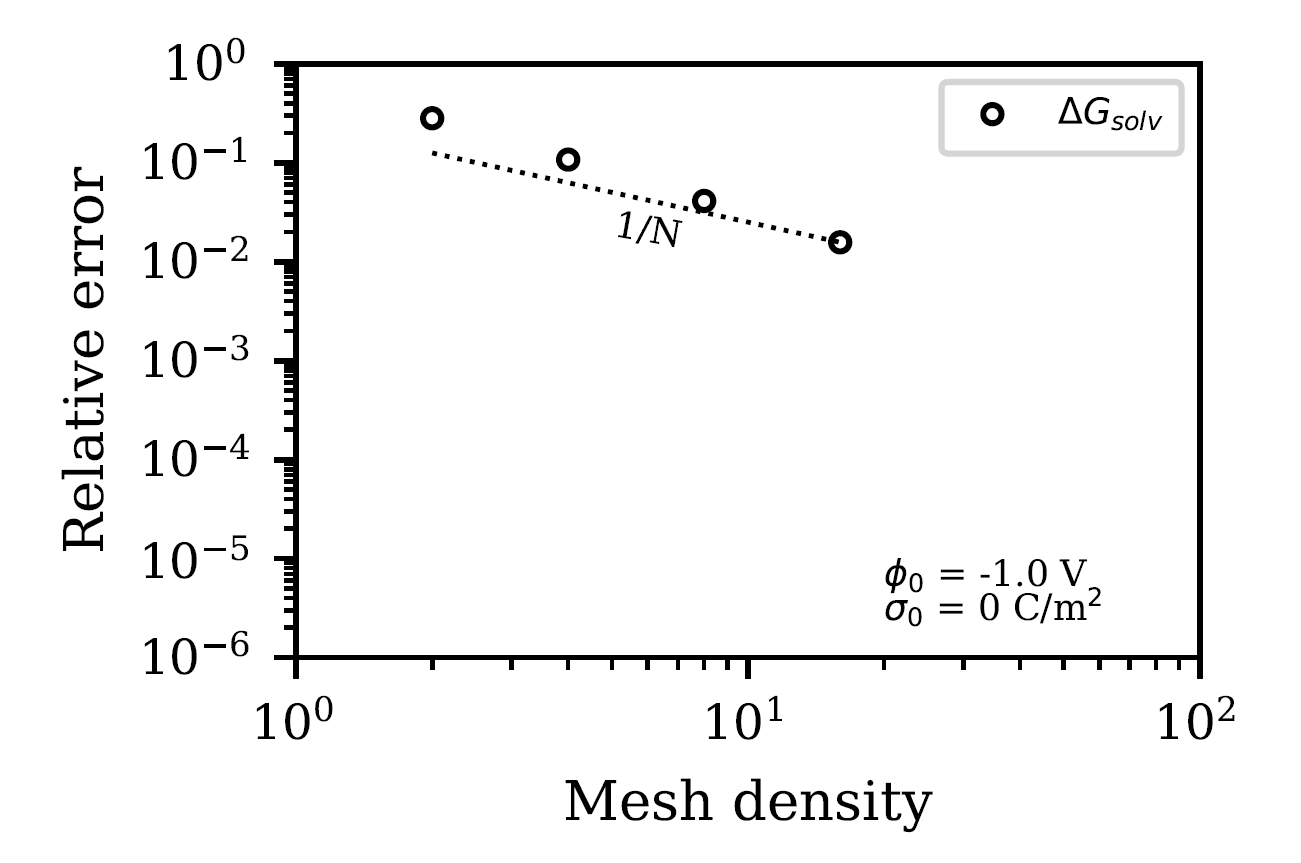}
    \includegraphics[width=0.45\textwidth]{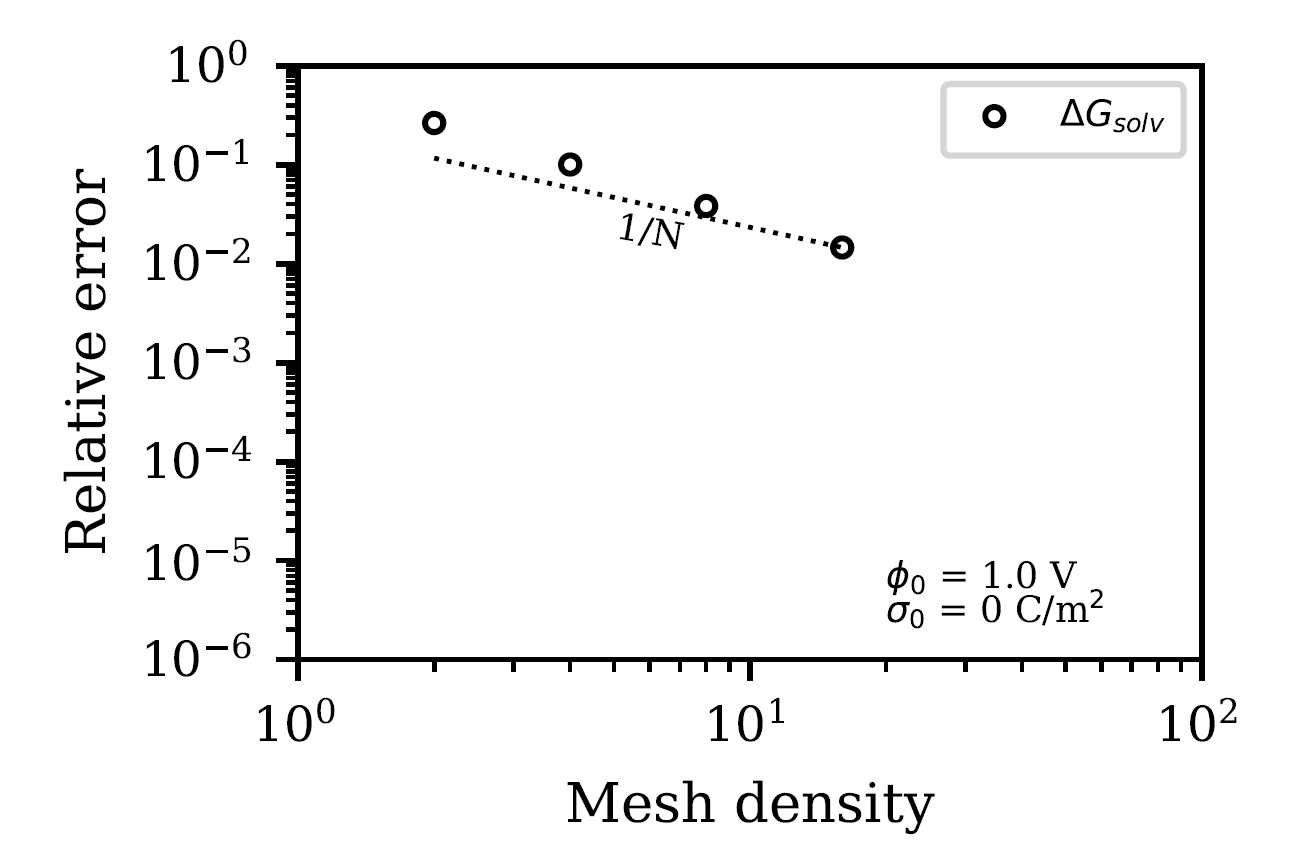}
    \caption{Error convergence as a function of grid size for cases with $\phi_0$=0 V and $\sigma_0$=0 C/m$^2$ (top left), $\phi_0$=-1.0 V and $\sigma_0$=0 C/m$^2$ (top right), and $\phi_0$=1.0 V and $\sigma_0$=0 C/m$^2$ (bottom).}
    \label{fig:mesh_refinement_1fni}
\end{figure}

We performed simulations with 2, 4, 8, and 16 elements per \AA$^2$ for $\alpha_{tilt} = 154^\circ$ and $\alpha_{rot} = 90^\circ$, and used Richardson extrapolation to obtain an approximate exact solution.~\cite{roache1998verification,CooperBardhanBarba2014} These values are detailed in Table \ref{table:trypsin_conver_data}. Figure \ref{fig:mesh_refinement_1fni} shows the error convergence with respect to the extrapolated value, with the expected 1/$N$ behavior.

\begin{table} 
\centering
\caption{PyGBe parameters for verification runs.}
\label{table:convergence_trypsin_param}
\begin{tabular}{ccccccc}
\hline
\multicolumn{3}{c}{\# Gauss points} & \multicolumn{3}{c}{Treecode} & GMRES     \\ \hline
$K$    & $K_{fine}$    & $N_{k}$    & $N_{crit}$  & P   & $\theta$ & tol.      \\ \hline
4      & 19            & 9          & 500         & 6  & 0.5      & $10^{-5}$ \\ \hline
\end{tabular}
\end{table}


\begin{table}
\centering
\caption{Approximate exact solution for solvation and interaction free energies with $\sigma_0$=0 C/m$^2$, using Richardson extrapolation for $\alpha_{tilt} = 154^\circ$ and $\alpha_{rot} = 90^\circ$.}
  \begin{tabular}{cccc}
    \hline
    Field   &$\Delta G^{extra}_{solv,int}$ & $\Delta G^{extra}_{solv,iso}$ & $\Delta G^{extra}_{int}$ \\
    V/\AA   & kcal/mol       &     kcal/mol          &      kcal/mol        \\
    \hline
    -1.0  &  -588.5230  &  -606.8187   &  18.2958  \\
    0.0     &  -606.6111 & -606.8187  & 0.2077  \\
    1.0     &   -624.7125  &  -606.8187   & -17.8936  \\
    \hline
  \end{tabular}
    
  \label{table:trypsin_conver_data}
\end{table}


\begin{table}[!h]
\centering
  \caption{Preferred orientation of trypsin adsorbed on a  surface with $\sigma_0$=0 C/m$^2$, under an electric field as a function of the salt concentration of the medium. These results are summarized by Figure 5.}
  \label{table:trypsin_kappa_1}
  \begin{tabular}{cccccc}
    \hline
     Field   & Ionic strength & $\kappa$ &  $\Delta G_{int}$ & $\alpha_{tilt}$ & $\alpha_{rot}$ \\
      V  &   mM   &  \AA$^{-1}$  & kcal/mol       &               &              \\
    \hline
                  & 10    & 0.032  & -20.3374 &  &  \\
    -1.0         & 50    & 0.0725 & -24.4049 & 48 & 320 \\
                  & 100   & 0.1026 & -21.7976 &  &  \\
                  & 150   & 0.125  & -18.7606 &  &  \\
    \hline 
                  & 10    & 0.032  & -24.7672 & 148 & 140 \\
     1.0        & 50    & 0.0725 & -37.0373 & 148 & 140 \\
                 & 100   & 0.1026 & -34.1979 & 144 & 160 \\
                 & 150   & 0.125  & -30.8499 & 144 & 160  \\
    
    \hline
  \end{tabular}
\end{table}

%

\bibliography{main}

%% file: abstract.tex
Under the most common experimental conditions, the adsorption of proteins to solid surfaces is a spontaneous process that leads to a rather compact layer of randomly oriented molecules. However, controlling such orientation is critically important for the development of catalytic surfaces. In this regard, the use of electric fields is one of the most promising alternatives. Our work is motivated by experimental observations that show important differences in catalytic activity of a trypsin-covered surface, which depended on the applied potential during the adsorption. Even though adsorption results from the combination of several physical processes, we were able to determine that (under the selected conditions) mean-field electrostatics dominates over other effects, determining the orientation and yielding a difference in catalytic activity. We simulated the electrostatic potential numerically, using an implicit-solvent model based on the linearized Poisson-Boltzmann equation. This was implemented in an extension of the code PyGBe that included an external electric field, and rendered the electrostatic component of the solvation free energy. Our model estimated the overall affinity of the protein with the surface, and their most likely orientation as a function of the potential applied. Our results show that the active sites of trypsin are, on average, more exposed when the electric field is negative, which agrees with the experimental results of catalytic activity, and confirms the premise that electrostatic interactions can be used to control the orientation of adsorbed proteins.

%% file: intro.tex
Controlling the orientation of proteins as they adsorb to surfaces is a challenging and critical endeavor. This issue is specially important for the development of biosensors and biocatalysts because the orientation of proteins with respect to the sorbent surface may affect the availability of the active sites \cite{doi:10.1021/la203095p,trilling2013antibody}. Since the adsorption process results from the combination of hydrophobic and electrostatic forces, it can be rationalized considering the physico-chemical properties of the surface, the protein’s 3D structure and amino acid sequence, as well as the experimental conditions. While hydrophobic substrates tend to interact with the hydrophobic core of the protein, driving conformational changes, hydrophilic surfaces tend to interact with the charged and polar functional groups of the protein’s surface, thus influencing their orientation \cite{latour2005biomaterials}. Along the same lines, recent studies have demonstrated that proteins appear to sense variations in the topography of their nanoscale environments, also potentially resulting in alterations of their orientation \cite{doi:10.1021/ja056278e}. In addition, several groups have developed approaches to influence the adsorption process (amount, orientation, kinetics, etc.) of proteins based on the natural heterogeneous distribution of charges \cite{doi:10.1021/la200141t, doi:10.1186/1559-4106-8-18}, by performing chemical modifications on either the -NH$_{2}$ or the -COOH terminal groups \cite{catal8050192, LEY2011539}, or by directed mutations\cite{doi:10.1021/ac101072h}.  Alternatively, and although it can only be applied to conductive materials, the adsorption process can also be influenced by the application of an external electric field \cite{BOS199491, moulton2003investigation, VanTassel2006, doi:10.1021/la4029657,BENAVIDEZ2014164}, potentially affecting the orientation \cite{emaminejad2015tunable, ghisellini2016direct, mulheran2016steering, martin2018electric, takahashi2018new}.

Given the importance of the orientation on the resulting bioactivity, the adsorption process has been also investigated with several computational models, ranging from highly detailed (but slow) molecular dynamics  \cite{liu2013multiscale,liu2015ribonuclease}, to more phenomenological (yet faster) coarse-grained approaches \cite{sheng2002orientation,zhou2003orientation,xie2010parallel}. The latter methods are usually based on bead-type descriptions, that agglomerate several atoms into a single effective bead. This comes at the cost of higher levels of parameterization. A possibility to overcome this limitation is to restrict the approximated representation to the solvent only, leaving the protein with an all-atom description. In this regard, implicit-solvent models are an interesting alternative because although they consider the solvent as a continuum dielectric medium, they provide a complete molecular description of the solute \cite{ROUX19991}. This sets a framework where continuum electrostatic theory can be easily applied. Under this premise, the ions present in the solvent are considered as point charges that are free to move in response to an electric field, and they arrange at equilibrium according to Boltzmann statistics. In turn, this gives rise to the Poisson-Boltzmann equation to solve for the electrostatic potential. The continuum description makes the calculation of free energies easy, and rather than studying the problem dynamically, a thermodynamic-state analysis becomes natural. Despite these advantages, two important drawbacks of this method remain: the assumption of a rigid solute and the need for well-parameterized dielectric constants. However, the implicit-solvent model remains an efficient approach to study the orientation of proteins at adsorption \cite{CooperClementiBarba2015,CooperBarba2016,malaspina2019protein}. While some computational studies have included the effect of external electric fields on the adsorption process, mainly with all-atom molecular dynamics \cite{xie2013effects, mulheran2016steering, xie2020molecular}, there is still a critical need to develop faster computational models to accurately describe and potentially predict how proteins would behave in the proximity to a surface charged via an external electric field.

Aiming to address this need and advance the development of bio-catalysts, this report describes an efficient computational model to predict the adsorption of proteins to conductive surfaces, as well as the possibility to affect their orientation using an external electric field. This model represents a new approach to describe adsorption under an external field using the Poisson-Boltzmann equation, enabling researchers to perform useful computations on workstations and small clusters, which are currently available in most universities and research centers around the world. In particular, the model extends the implicit solvent Poisson-Boltzmann model implemented in PyGBe \cite{CooperBardhanBarba2014,cooper2016pygbe}, to include an external electric field. PyGBe is a boundary-element solver, that computes the electrostatic component of the solvation free energy. Here, this free energy is used to calculate the probability of different orientations. Similar computational models have been used to study ellipsoidal colloids near charged surfaces under an external field \cite{tsori2020bistable}, however, in our case we need to include the structural information of the adsorbed molecule.

Due to its industrial and clinical importance, trypsin was the model protein of choice to measure the relevance of electrostatics in the orientation, and validate the outcomes of the model. To this end, the catalytic activity of trypsin (adsorbed on carbon electrodes at different potentials) serves as an experimental approach to evaluate the capabilities of the model.

%% file: methods.tex
\subsection{Mathematical model}

\subsubsection{The Poisson-Boltzmann equation with a boundary element method (BEM)}

\begin{figure}[h]
\centering
  \includegraphics[width=0.6\textwidth]{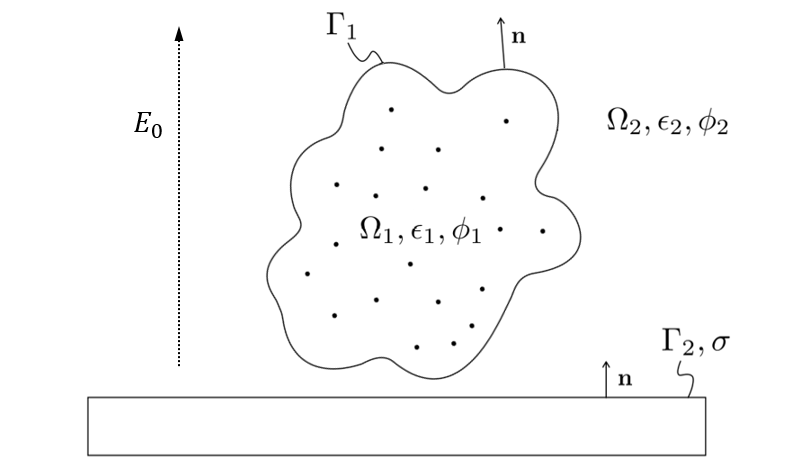}
  \caption{Sketch of a molecule interacting with a charged surface under an external electric field: $\Omega_{1}$ is the protein, $\Omega_{2}$ is the solvent, $\Gamma_{1}$ is the SES, and $\Gamma_{2}$ the surface with a prescribed charged density.}
  \label{fig:Protein-Surface_sketch}
\end{figure}

The implicit solvent model considers an infinite {\it solvent} domain that contains salt and has a high permittivity, with a low-dielectric cavity: the {\it solute} domain \cite{ROUX19991,decherchi2015implicit}. These two regions are interfaced by the molecular surface, in this case, the solvent-excluded surface~\cite{Connolly83}. Applying continuum electrostatic theory gives rise to a system of partial differential equations for the potential, where the Poisson-Boltzmann (solvent region) and Poisson (solute region) equations are coupled through continuity conditions on the molecular surface. This model can be extended to consider surfaces with imposed charge or potential \cite{CooperClementiBarba2015,CooperBarba2016}, or external electric fields \cite{roux1997influence}. The mathematical formulation to model the setup shown in Fig. \ref{fig:Protein-Surface_sketch} with a charged surface and an external field, is
\begin{align}\label{eqn:ProtySurfModel}
\nabla^{2}\phi_{1}(\mathbf{r}) = - \sum_{k}^{Nc} \dfrac{q_{k}}{\epsilon_{1}}\delta(\mathbf{r},\mathbf{r}_{k}) & \quad \mathbf{r}\in  \Omega_{1}\nonumber\\
\nabla^{2}\phi_{2}(\mathbf{r}) = \kappa^{2} \phi_{2}(\mathbf{r})& \quad \mathbf{r}\in  \Omega_{2} \nonumber \\
\phi_{1} = \phi_{2} + \phi_{e} & \quad \mathbf{r}\in  \Gamma_{1} \nonumber\\
\epsilon_{1} \dfrac{\partial \phi_{1}}{\partial \mathbf{n}} = \epsilon_{2}\left(\dfrac{\partial \phi_{2}}{\partial \mathbf{n}} + \dfrac{\partial \phi_{e}}{\partial \mathbf{n}}\right) &\nonumber\\
 -\epsilon_{2}\left(\dfrac{\partial \phi_{2}}{\partial \mathbf{n}} + \dfrac{\partial \phi_{e}}{\partial \mathbf{n}}\right)= \sigma_{0} &\quad \mathbf{r}\in \Gamma_{2},
\end{align}

where $\mathbf{n}$ is a unit vector normal to the surfaces $\Gamma_1$ and $\Gamma_2$, pointing into the solvent, $\sigma_0$ is the surface charge on $\Gamma_2$, and $\epsilon_1$ and $\epsilon_2$ are the dielectric constants in the corresponding regions. The total electrostatic potential in $\Omega_1$ is $\phi_1$, whereas in $\Omega_2$ this is decomposed into $\phi_e=\phi_0e^{-\kappa z}$, due to the external field, and the remaining $\phi_2$.

Using Green's second identity on Eq. \eqref{eqn:ProtySurfModel}, we can write $\phi_1$ and $\phi_2$ in terms of boundary integral equations as
\begin{align}\label{eq:bie}
\phi_{1}(\mathbf{r}) &= - K_{\mathbf{r},L}^{\Gamma_1}\left(\phi_{1, \Gamma_1}\right) + V_{\Gamma_{1},L}^{\mathbf{r}}\left(\dfrac{\partial}{\partial \mathbf{n}} \phi_{1, \Gamma_1}\right) +  \dfrac{1}{\epsilon_{1}} \sum_{k}^{N_{c}} \dfrac{q_{k}}{4\pi |\mathbf{r} - \mathbf{r}_{k}|} \quad \mathbf{r}\in\Omega_1 \nonumber\\
\phi_{2}(\mathbf{r}) &= K_{\Gamma_{1},Y}^{\mathbf{r}}(\phi_{2, \Gamma_1}) - V_{\Gamma_{1},Y}^{\mathbf{r}}\left(\dfrac{\partial}{\partial \mathbf{n}} \phi_{2, \Gamma_1}\right) + K_{\Gamma_{2},Y}^{\mathbf{r}}(\phi_{2, \Gamma_2}) - V_{\Gamma_{2},Y}^{\mathbf{r}}\left(\dfrac{\partial}{\partial \mathbf{n}} \phi_{2, \Gamma_2}\right) \quad \mathbf{r}\in\Omega_2   
\end{align}
where,
\begin{align}
V_{\Gamma_{a}}^{\mathbf{r}}\left(\varphi\right) &= \oint_{\Gamma_a} G(\mathbf{r},\mathbf{r}')\varphi(\mathbf{r}')\rm{d}\mathbf{r}'\nonumber\\
K_{\Gamma_{a}}^{\mathbf{r}}\left(\varphi\right) &= \oint_{\Gamma_a} \frac{\partial G}{\partial\mathbf{n}}(\mathbf{r},\mathbf{r}')\varphi(\mathbf{r}')\rm{d}\mathbf{r}'
\end{align}
are the single and double layer potentials of a distribution $\varphi(\mathbf{r}_a)$ on $\Gamma_a$, evaluated at a point $\mathbf{r}$ located anywhere in the domain, except $\Gamma_a$. $G(\mathbf{r},\mathbf{r}')$ is the Green's functions of the Poisson or linearized Poisson-Boltzmann equations. These expressions are also known as the Laplace and Yukawa potentials, respectively:
\begin{align}
G_L(\mathbf{r},\mathbf{r}') &= \frac{1}{4\pi|\mathbf{r}-\mathbf{r}'|} \nonumber\\
G_Y(\mathbf{r},\mathbf{r}') &= \frac{e^{-\kappa|\mathbf{r}-\mathbf{r}'|}}{4\pi|\mathbf{r}-\mathbf{r}'|} 
\end{align}
with $\kappa$ the inverse of the Debye length. We can build a system of equations to compute the potential and its normal derivative on the surface by taking the limit as $\mathbf{r}\to\Gamma_1$ and $\mathbf{r}\to\Gamma_2$, which leaves 
\begin{eqnarray}\label{eq:bie_limit}
\dfrac{\phi_{1, \Gamma_1}}{2} + K_{\Gamma_{1},L}^{\Gamma_1}\left(\phi_{1, \Gamma_1}\right) - V_{\Gamma_{1},L}^{\Gamma_1}\left(\dfrac{\partial}{\partial n} \phi_{1, \Gamma_1}\right) = \dfrac{1}{\epsilon_{1}} \sum_{k}^{N_{c}} \dfrac{q_{k}}{4\pi |r_{\Gamma_1} - r_{k}|} \nonumber\\
\dfrac{\phi_{2, \Gamma_1}}{2} - K_{\Gamma_{1},Y}^{\Gamma_1}(\phi_{2, \Gamma_1}) + V_{\Gamma_{1},Y}^{\Gamma_1}\left(\dfrac{\partial}{\partial n} \phi_{2, \Gamma_1}\right) - K_{\Gamma_{2},Y}^{\Gamma_1}(\phi_{2, \Gamma_2}) + V_{\Gamma_{2},Y}^{\Gamma_1}\left(\dfrac{\partial}{\partial n} \phi_{2, \Gamma_2}\right) = 0\nonumber\\
\dfrac{\phi_{2, \Gamma_2}}{2} - K_{\Gamma_{1},Y}^{\Gamma_2}(\phi_{2, \Gamma_1}) + V_{\Gamma_{1},Y}^{\Gamma_2}\left(\dfrac{\partial}{\partial n} \phi_{2, \Gamma_1}\right) - K_{\Gamma_{2},Y}^{\Gamma_2}(\phi_{2, \Gamma_2}) + V_{\Gamma_{2},Y}^{\Gamma_2}\left(\dfrac{\partial}{\partial n} \phi_{2, \Gamma_2}\right) = 0 
\end{eqnarray}
where the subscripts on $V$ and $K$ indicate the surface where $\mathbf{r}$ is evaluated. Then, we apply the interface conditions from Eq. \eqref{eqn:ProtySurfModel}, to write Eq. \eqref{eq:bie_limit} in matrix form as
\begin{eqnarray}\label{eq:matrix_form}
  \begin{bmatrix}
  1/2 + K_{\Gamma_{1},L}^{\Gamma_{1}} & - V_{\Gamma_{1},L}^{\Gamma_{1}} & 0 \\
  1/2 - K_{\Gamma_{1},Y}^{\Gamma_{1}} &  \dfrac{\epsilon_{1}}{\epsilon_{2}}V_{\Gamma_{1},Y}^{\Gamma_{1}} & - K_{\Gamma_{2},Y}^{\Gamma_{1}} \\
  - K_{\Gamma_{1},Y}^{\Gamma_{2}} & \dfrac{\epsilon_{1}}{\epsilon_{2}}V_{\Gamma_{1},Y}^{\Gamma_{2}} & (1/2 - K_{\Gamma_{2},Y}^{\Gamma_{2}}) 
 \end{bmatrix}
 \begin{bmatrix}
  \phi_{1}(r_{\Gamma_{1}}) \\
  \dfrac{\partial}{\partial n} \phi_{1}(r_{\Gamma_{1}})  \\
  \phi_{2}(r_{\Gamma_{2}})  \\
 \end{bmatrix}
    = \nonumber\\
 \begin{bmatrix}
  \dfrac{1}{\epsilon_{1}} \sum_{k}^{N_{c}} \dfrac{q_{k}}{4\pi |r_{\Gamma_{1}} - r_{k}|}\\
  (1/2 - K_{\Gamma_1,Y}^{\Gamma_1})\phi_{e,\Gamma_1} + V_{\Gamma_1,Y}^{\Gamma_1}\dfrac{\partial}{\partial n} \phi_{e,\Gamma_1} + V_{\Gamma_{2},Y}^{\Gamma_{1}}\left(\dfrac{\sigma_{0}}{\epsilon_{2}} + \dfrac{\partial}{\partial n} \phi_{e,\Gamma_2}\right)  \\
 - K_{\Gamma_1,Y}^{\Gamma_2}\phi_{e,\Gamma_1} + V_{\Gamma_1,Y}^{\Gamma_2}\dfrac{\partial}{\partial n} \phi_{e,\Gamma_1} + V_{\Gamma_{2},Y}^{\Gamma_{2}}\left(\dfrac{\sigma_{0}}{\epsilon_{2}} + \dfrac{\partial}{\partial n} \phi_{e,\Gamma_2}\right)\\
 \end{bmatrix}
\end{eqnarray}

We solved the linear system in Eq.\eqref{eq:matrix_form} to obtain the electrostatic potential on the surface with the GMRES \cite{saad1986gmres} solver implemented in the boundary element method software PyGBe \cite{CooperBardhanBarba2014,cooper2016pygbe}. This code considered a piecewise constant distribution of the potential and its normal derivative on the triangulated molecular surface, and used centroid collocation. In this work, we extended the model in PyGBe to account for an external electric field. These extensions were performed on a fork\footnote{\url{https://github.com/UrzuaSergio/ElectricFieldPyGBe}} of the official GitHub\footnote{\url{https://github.com/pygbe}} repository of the code.

PyGBe allowed solving the integrals on the surface with Gauss quadrature rules depending on the distance between the collocation point and the boundary element. In particular, the code used $K$ points if they are far away, $K_{fine}$ if they are close by, and a semi-analytical technique \cite{hess1967calculation} with $N_k$ points on each edge of the triangle if the integral is singular. In this context, a matrix-vector product became an N-body problem, where Gauss quadrature nodes serve as sources of mass, and collocation points as evaluation centers. PyGBe accelerated each vector-product in the GMRES solver with a treecode algorithm \cite{barnes1986hierarchical,duan2001adaptive,li2009cartesian}. The treecode groups the Gauss points in a tree structure, and approximates far-field interactions between a box and a collocation point with a Taylor series expansion of order $P$. The multipole-acceptance criterion $\theta>\frac{R_b}{R}$ defined if a box is far enough, where $R_b$ and $R$ are the box size and the distance between the collocation point and the cluster of Gauss nodes, respectively. 

\subsubsection{Energy calculation}

We were interested in computing the interaction free energy of specific conformations of trypsin near the charged surface, to determine the probability of each occurrence. The interaction free energy is the difference in free energy between the setup with trypsin and the charged surface interacting under the external electric fields ($\Delta G_{total}^{sys}$), and each entity  isolated without an external field ($\Delta G_{total}^{tryp}$ and $\Delta G_{total}^{surf}$). This is
\begin{equation}\label{eq:Gint}
\Delta G_{int} = G_{total}^{sys} - G_{total}^{tryp} - G_{total}^{surf}.
\end{equation}
The total free energy $G_{total}$ has three sources: solvation ($\Delta G_{solv}$), surface ($G_{surf}$), and Coulomb ($G_{coul}$). In the first place, the solvation free energy is the free energy difference of a system in vacuum and dissolved states. For the solute molecule, this can be computed as \cite{che2008electrostatic} 
\begin{equation}\label{eq:dGsolv}
    \Delta G_{solv}=\dfrac{1}{2}\int_{\Omega}\rho\phi_{react} =\frac{1}{2}\sum_{k=0}^{N_{q}} q_{k}\phi_{react}(r_{k})
\end{equation}
where $\phi_{reac}$ is the change in electrostatic potential as the solute is placed inside the solvent. If we subtract the Coulomb potential out of Eq. \eqref{eq:bie}, $\phi_{reac}$ appears as 
\begin{equation}\label{eq:phi_reac}
    \phi_{reac}(\mathbf{r}) = - K_{\mathbf{r},L}^{\Gamma_1}\left(\phi_{1, \Gamma_1}\right) + V_{\Gamma_{1},L}^{\mathbf{r}}\left(\dfrac{\partial}{\partial \mathbf{n}} \phi_{1, \Gamma_1}\right).
\end{equation}

Secondly, the surface free energy ($G_{surf}$) is calculated with \cite{carnie1993interaction}
\begin{equation}\label{eq:Gsurf}
G_{surf}= \dfrac{1}{2}\int_{\Gamma}(\phi_2+\phi_e)\sigma_{0}d\Gamma = \dfrac{1}{2}\sum_{j=1}^{N_{p}}(\phi_2(r_{j})+\phi_e(r_{j}))\sigma_{0_{j}}A_{j}
\end{equation}
where the sum is over all $N_p$ boundary elements on the charged surface. 

Finally, as the Coulomb energy does not change as the molecule is dissolved, it cancels out of the calculation.

\subsubsection{Sampling orientations}

By having the interaction free energy for every orientation from Eq.~\eqref{eq:Gint}, we can determine its probability of occurrence using Boltzmann statistics. In particular, the probability of finding the system in a state $\lambda$ is
\begin{equation}\label{eq:prob_mol_I}
    P(\lambda)=\dfrac{\int_{\lambda}\exp\left(-\dfrac{\Delta G_{int}}{k_{B}T}\right) d\gamma}{\int_{\Lambda} \exp\left(-\dfrac{\Delta G_{int}}{k_{B}T}\right)d\gamma}
\end{equation}
where $\Lambda$ considers all possible states, $k_B$ is the Boltzmann constant, and $T$ is the temperature. We sample all possible orientations by aligning the dipole moment vector of trypsin to the normal vector to the surface, and tilting trypsin an angle $\alpha_{tilt}$ that varies from 0$^\circ$ to 180$^\circ$. Then, we rotate about the dipole moment vector an angle $\alpha_{rot}$ from 0$^\circ$ to 360$^\circ$, to obtain every possible orientation. This way, we can compute the integrals in Eq. \eqref{eq:prob_mol_I} as 
\begin{equation}\label{eqn:prob_mol_II}
\int_{\lambda} \exp\left(-\dfrac{\Delta G_{int}}{k_{B}T}\right) d\gamma=\int \int \exp\left(-\dfrac{\Delta G_{int}}{k_{B}T}\right)d\alpha_{rot}d\alpha_{tilt}
\end{equation}
where the angles $\alpha_{tilt}$ and $\alpha_{rot}$ define the state $\lambda$. This process considers a small constant distance between trypsin and the surface of 2\AA, as we are studying the orientation at adsorption. 

\subsubsection{Quantifying accessibility to active sites}

The accessibility of antigens to binding sites changes for different orientations, affecting the catalytic activity of the surface. To quantify the accessibility, we defined a factor $f = z_i/z_{max}$, where $z_i$ is the distance between the binding site and the electrode surface, and $z_{max}$ is the maximum $z$ position of the atoms in that same orientation, making $0 < f \leq 1$. Even though there are situations where an active site with a high $f$ may be facing into the surface, and hence, not accessible, $f$ still is a good indicator of accessibility. Using the probability from Eq. \eqref{eq:prob_mol_I}, we can define an average accessibility as
\begin{equation}\label{eq:avg_access}
    \bar{f} = \int_\Lambda P(\lambda)f(\lambda) d\lambda
\end{equation}

\subsection{Experimental details}
\subsubsection{Reagents} 
We used sodium bicarbonate, trypsin from porcine pancreas (T4799), and a trypsin activity kit (MAK290) purchased from Sigma-Aldrich (St. Louis, MO, USA). The aqueous solutions were prepared using 18 M$\Omega\cdot$cm water (NANOpure Diamond, Barnstead; Dubuque, IA) and analytical reagent grade chemicals. The phosphate buffer solution was prepared by dissolving anhydrous Na$_{2}$HPO$_{4}$ (Fisher Scientific; Fair Lawn, NJ, USA) in ultrapure water. We measured the pH of the solutions using a glass electrode connected to a digital pH meter (Orion 420A+, Thermo; Waltham, MA, USA) and adjusted with 0.1M solutions of HCl.

\subsubsection{Fabrication of carbon electrodes} 
Following the procedure described in previous publications from our group \cite{gomez2019co, giuliani2016},  we used electrodes obtained by pyrolysis of paper strips (4.5 cm × 1.5 cm; Whatman 3MM chromatography paper; GE Health Care; Pittsburgh, PA) using a tube furnace (Type F21100, Barnstead–Thermolyne; Dubuque, IA, USA). The quartz tube was first flushed with forming gas ( 5\% H$_{2}$ / 95\% Ar, 1 L/min) for 5 min (to remove the O$_{2}$ and avoid oxidation reactions) and then allowed to reach a temperature of 1000$^\circ$C, at a rate of 20$^\circ$C min/1. After 1 h, we turned off the tube furnace and allowed to cool-down to room temperature while maintaining the flow of forming gas. Finally, the pyrolyzed samples were removed from the furnace and stored in a Petri dish until use. The pyrolyzed paper layers were fixed to a Plexiglas substrate with double-sided tape and cut using a commercial 30W CO$_{2}$ laser engraver (Mini24, Epilog Laser Systems; Golden, CO, USA). The resulting electrodes featured a circular pad (where the reaction takes place, 0.50 cm$^{2}$) and a stem, similar to those previously reported \cite{giuliani2016}. Then, we applied silver paint (SPI Supplies; West Chester, PA, USA) to improve the electrical connection with the alligator clip connected to the potentiostat. To prevent water from wicking up the stem of the electrode (and increasing the electrode area), we applied parafilm to base of the stem, between the circular pad and the contact area.

\subsubsection{Electrochemical Techniques} 
We performed cyclic voltammetry (CV) on all electrodes to verify their functionality and electrical connections. In all cases, a standard three-electrode cell comprised of the carbon electrode, a silver/silver chloride (Ag/AgCl/KCl$_\text{sat}$), and a platinum wire were used as working, reference, and counter electrodes, respectively. The CV experiments used a CHI660A Electrochemical Analyzer (CH Instruments, Inc.; Austin, TX). For the adsorption of the trypsin under the application of a selected potential, we used the amperometry mode. For these cases, the electrode was immersed for 15 min in a solution containing 1 mg/mL of trypsin dissolved in phosphate buffer (0.1M, pH=8), with periodic mixing. These electrodes were then removed from the cell, thoroughly rinsed with DI water to remove any unbound protein from the surface, and immediately assayed for activity. 

\subsubsection{Trypsin Activity Assays} 
The activity of the adsorbed layer of trypsin was assessed colorimetrically following the cleavage of a substrate to generate p-nitroaniline (p-NA). Considering that the commercial kit is designed to accommodate small liquid samples placed on a microplate reader, the procedure was slightly modified. Specifically, 2 $\mu$L of the substrate were mixed with 98 $\mu$L of the buffer provided and with 400 $\mu$L of DI water in a standard eppendorf tube. The latter was added to ensure that, upon insertion, the electrode remained immersed and in direct contact with the solution containing the substrate. After two hours at room temperature, the electrode was removed from the tube, the solution centrifuged (10,000 rpm for 3 min) and then measured using an spectrophotometer (GENESYS$^\text{TM}$ 10 Series, Thermo; Madison, WI) at 405 nm.

%% file: results.tex
\subsection{Catalytic activity of trypsin adsorbed under different applied potentials. }

\begin{figure}
\centering
  \includegraphics[width=0.75\textwidth]{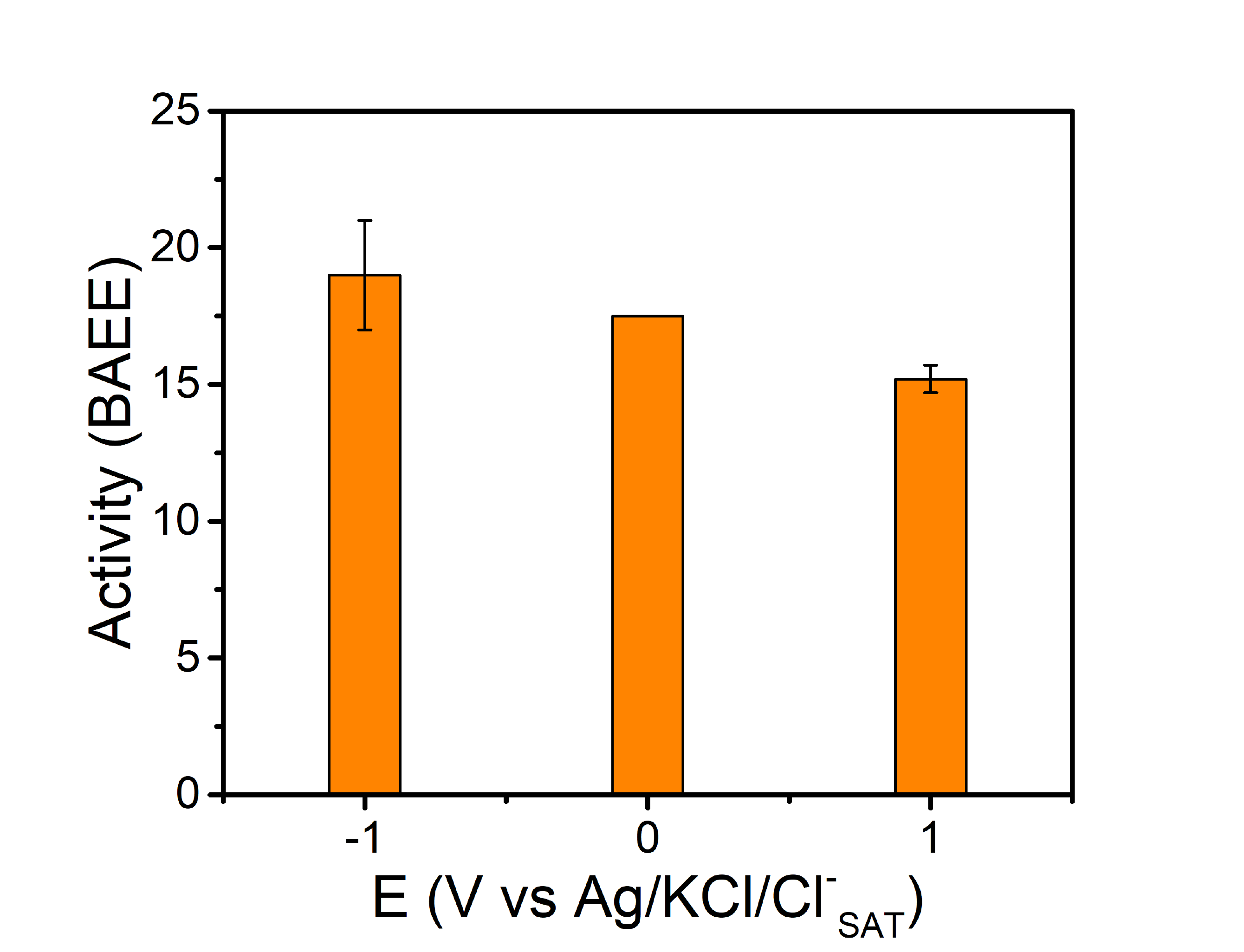}
  \caption{Enzymatic activity of the carbon electrodes modified by adsorbing trypsin at different potentials. Electrodes were immersed for 15 min in a solution containing 1 mg/mL trypsin dissolved in 0.1 M phosphate buffer, pH = 8}
  \label{fig:figactivity}
\end{figure}

In order to determine the effect of the potential applied to the electrode on the adsorption process, we measured the activity of trypsin immediately after the adsorption step. It is important to mention that these results involved an adsorption step of only 15 min, selected (along pertinent experimental conditions) to minimize the possibility to induce polarization of the protein and the subsequent formation of multilayers \cite{BENAVIDEZ2014164,doi:10.1021/la504890v} that could lead to a difference in adsorbed amount. Thus, we evaluated the catalytic activity of the resulting substrate by following the cleavage of a substrate to generate p-nitroaniline (which absorbs light at 405 nm). The results are summarized in Figure \ref{fig:figactivity}, where the resulting enzymatic activity is presented as a function of the potential applied during the adsorption stage. As it can be observed, all substrates showed significant activity of the adsorbed enzyme, where the highest (19 $\pm$ 2 U) and lowest (15.2 $\pm$ 0.5 U) activity correspond to adsorption at potentials of -1.0V or +1.0V, respectively. As expected, a control experiment, performed by adsorbing the enzyme at open circuit potential, rendered an intermediate activity (17.5 U).

The differences observed in Fig. \ref{fig:figactivity}, which we can associate to the effect of the electric field on the orientation of the adsorbed enzyme, served as motivation to develop the computational model.

\subsection{Simulations of trypsin adsorbing on an uncharged surface under an external potential}

To understand the role of electrostatics in the catalytic activity difference described in Fig. \ref{fig:figactivity}, we performed Poisson-Boltzmann simulations of trypsin (PDB code 1FNI) adsorbed on a surface with $\sigma_0$=0 C/m$^2$ (placed at a distance of 2 \AA) with an external field, and studied its orientation. In these calculations, the external potential was set to $\phi_0=\pm$1 V on the electrode surface. Likewise, permittivity values of $\epsilon_2$=80 (solvent) and $\epsilon=4$ (solute), and an inverse of Debye length $\kappa$=0.175~\AA$^{-1}$, valid for 0.1 M of phosphate buffer, were considered.

The Supplementary Information (Table S3) includes a careful mesh convergence analysis. In this case, we used a mesh density of 8 elements per \AA$^2$ on the molecular surface, and 2 on the electrode. The code parameters for PyGBe are presented by Table \ref{table:orientation_trypsin_param}. With these settings, the interaction energies were at most 1.68\% away from the extrapolated values shown in Table S3.  

\begin{figure}
    \centering
    \includegraphics[width=0.4\textwidth]{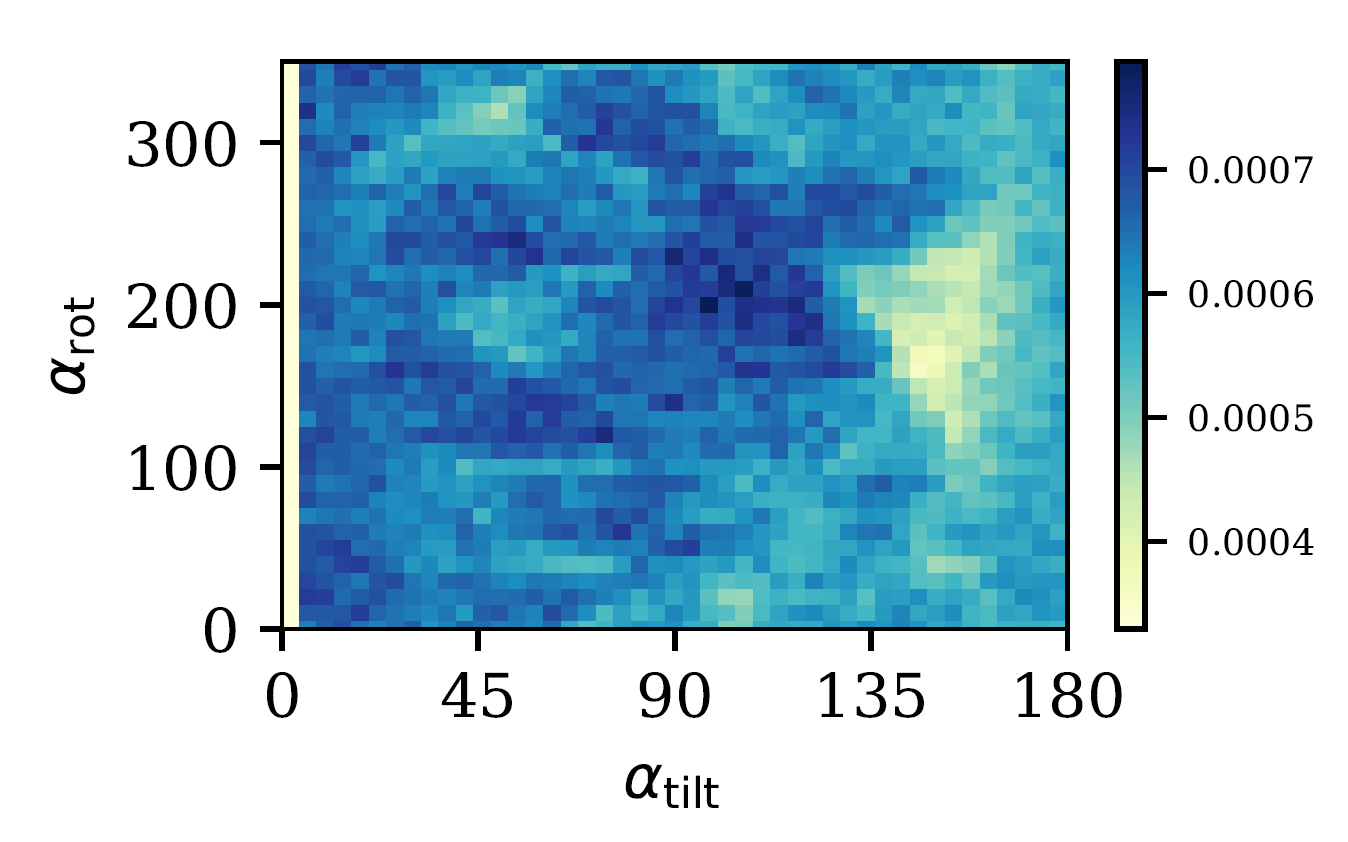}
    \includegraphics[width=0.4\textwidth]{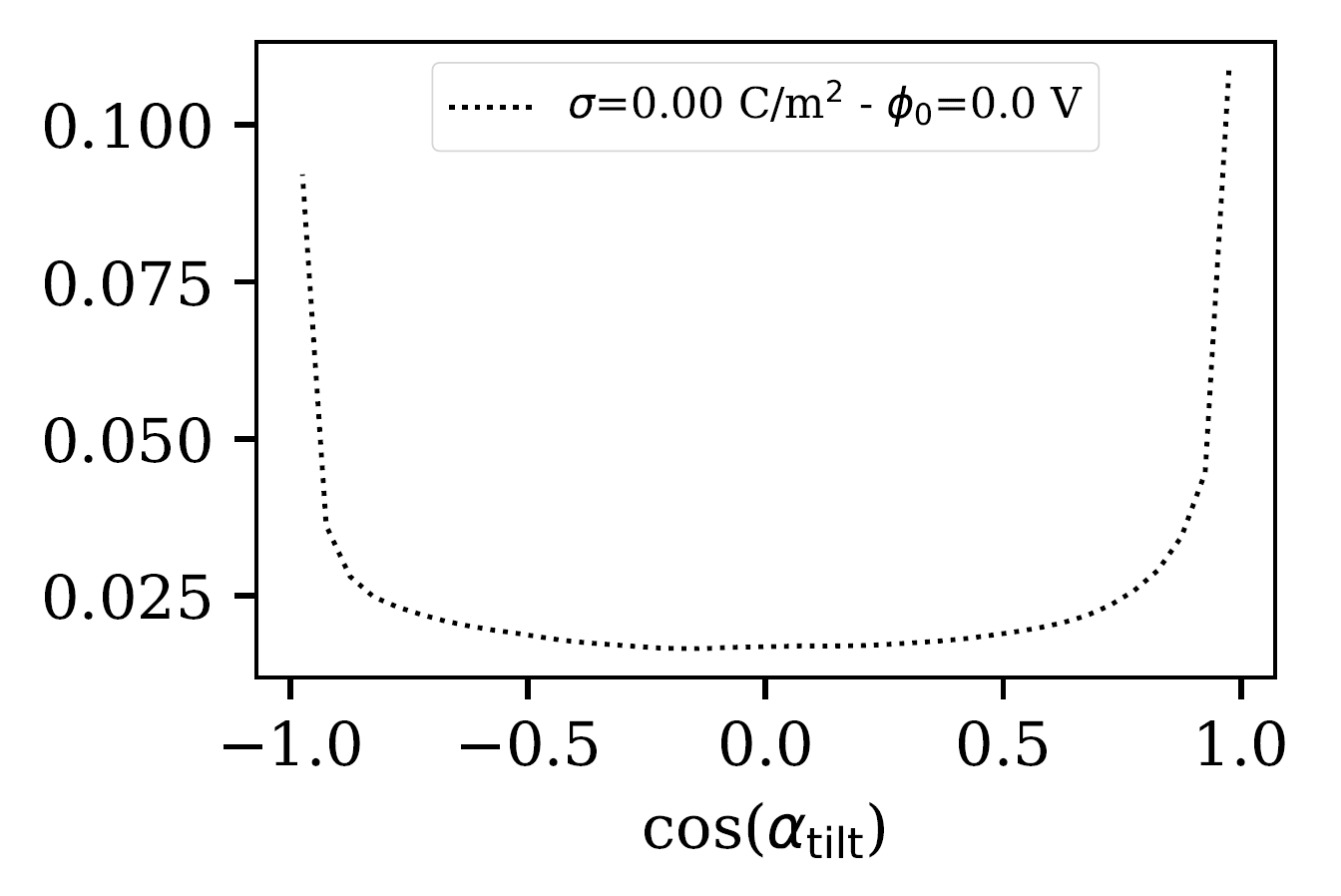}
    \includegraphics[width=0.4\textwidth]{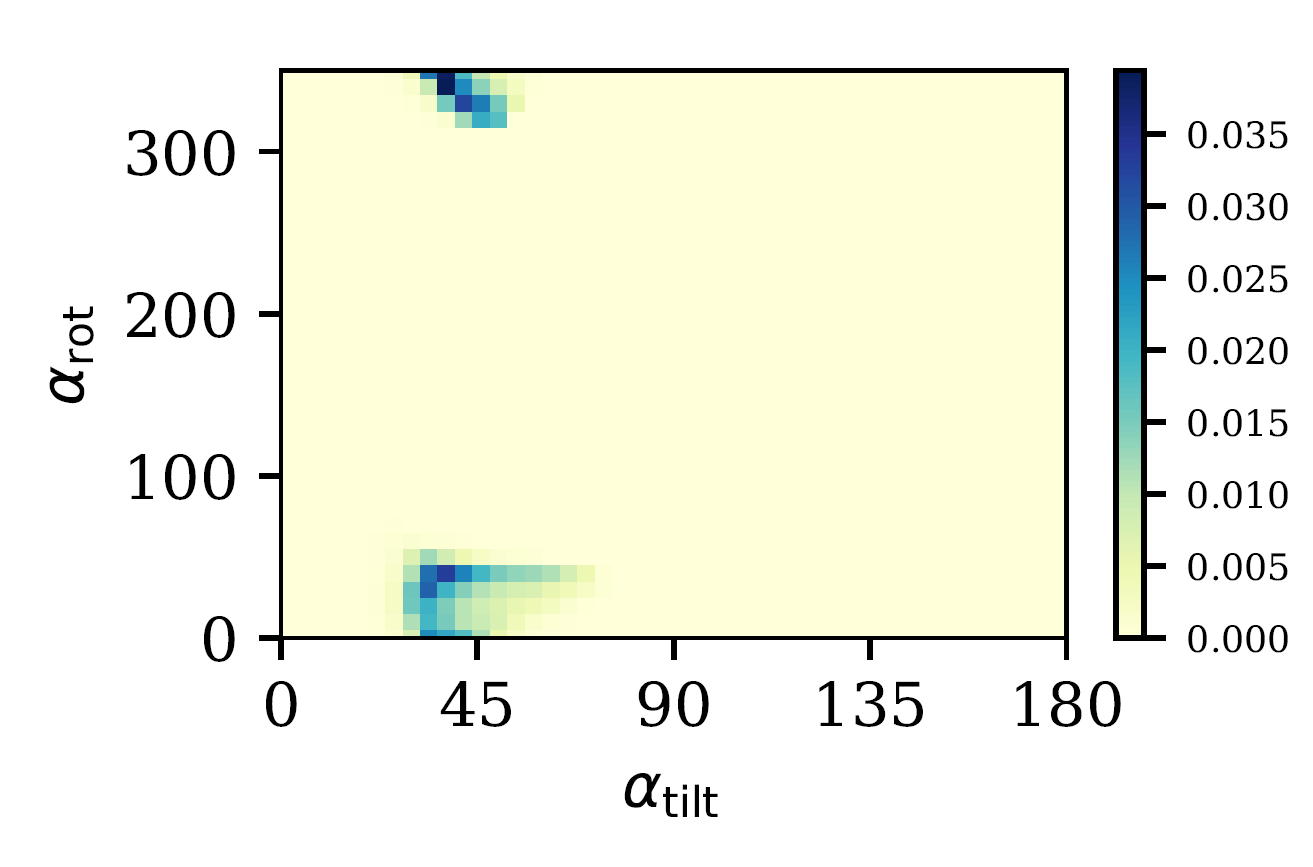}
    \includegraphics[width=0.4\textwidth]{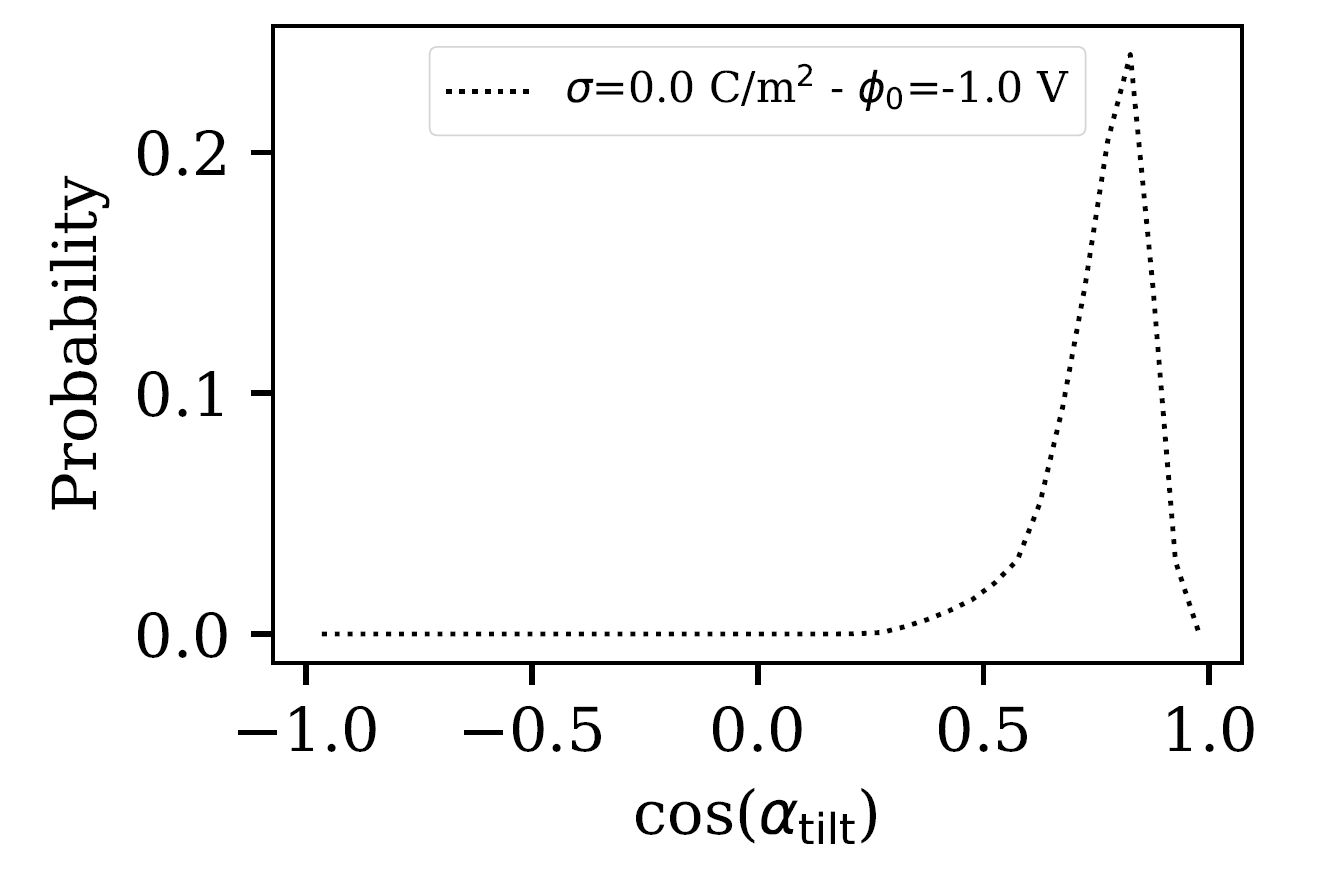}
    \includegraphics[width=0.4\textwidth]{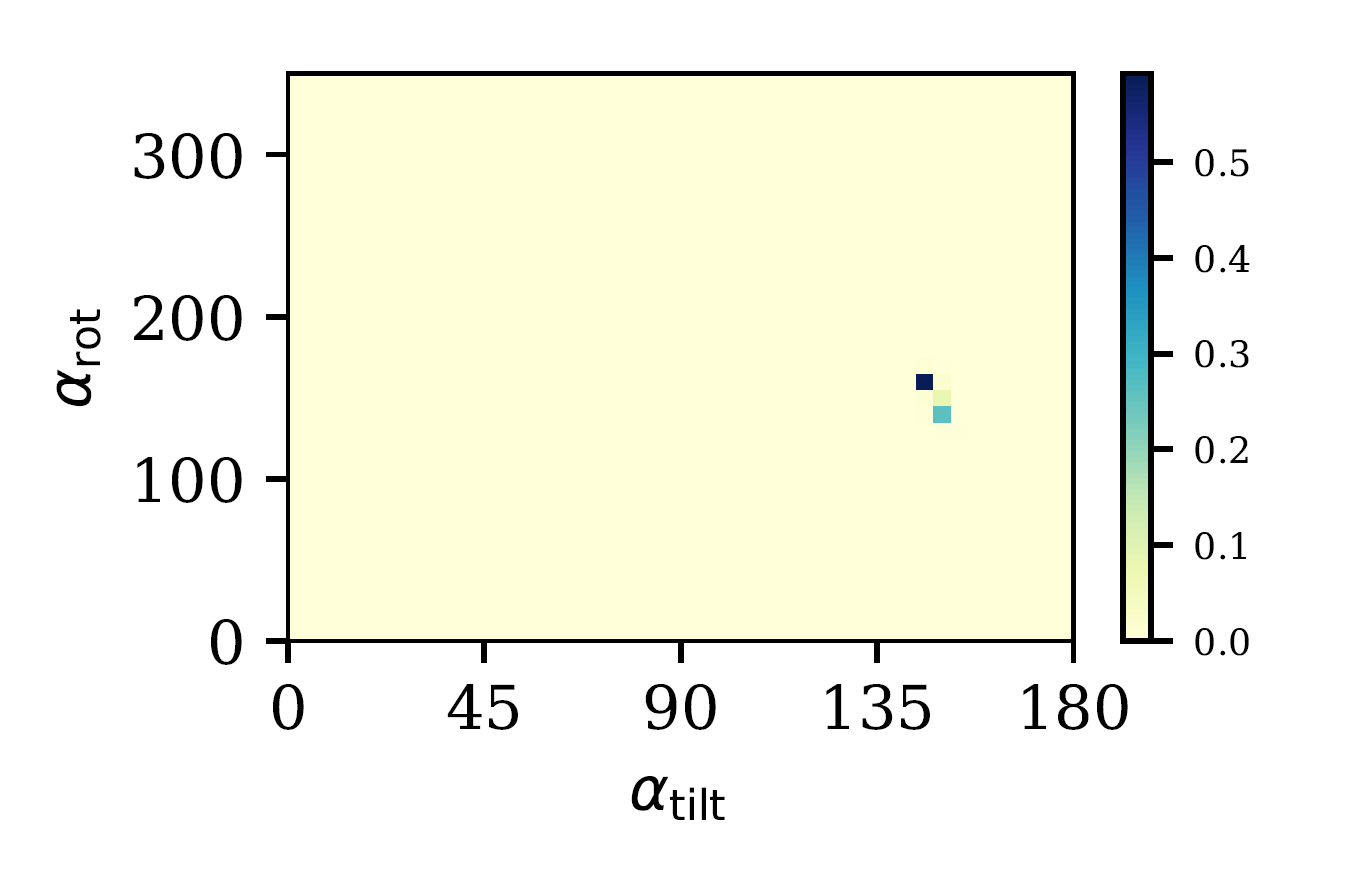}
    \includegraphics[width=0.4\textwidth]{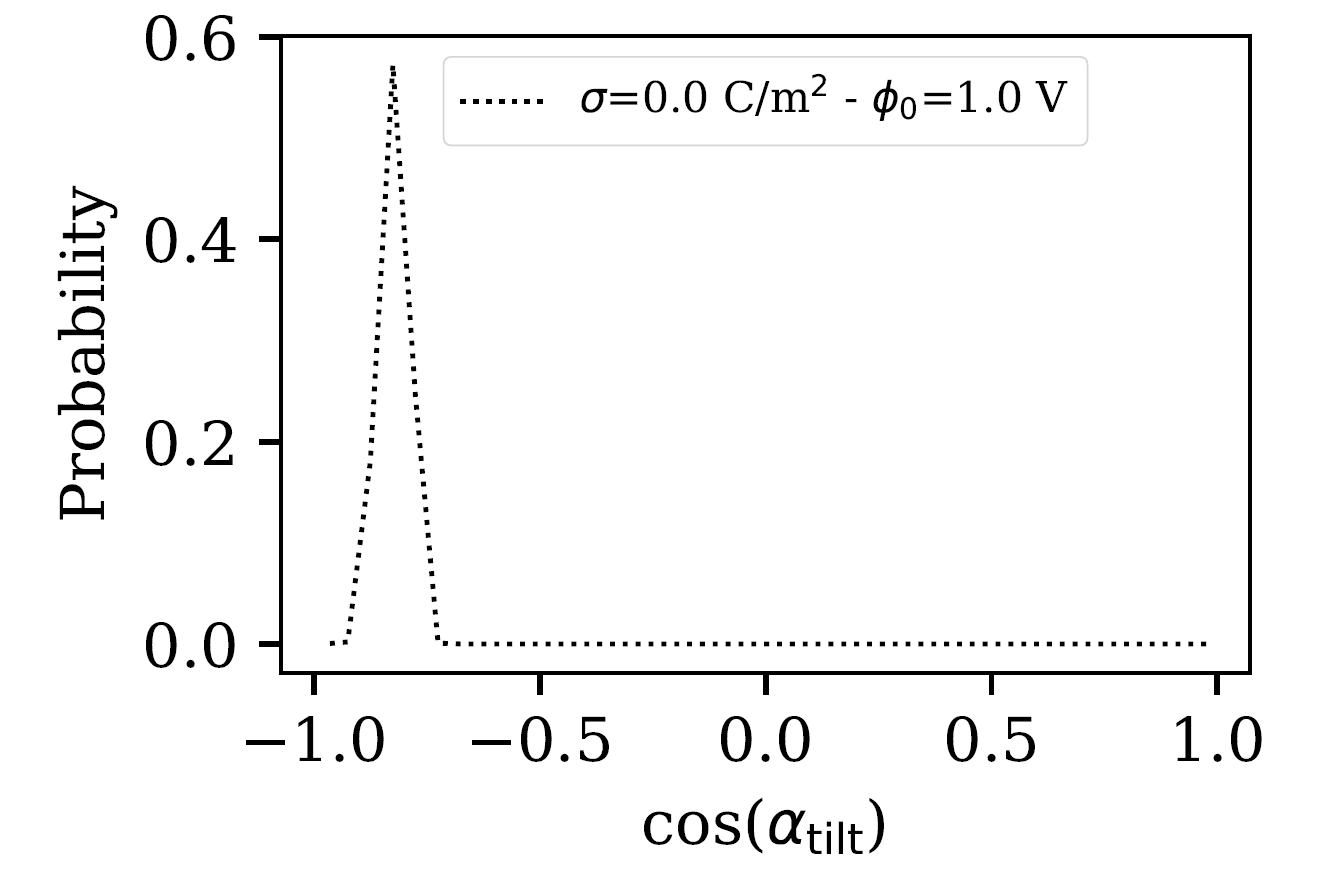}
    \caption{Orientation distribution for $\phi_0$=0 (top) $\phi_0$=-1 V (middle) and $\phi_0$=1 V (bottom).}
    \label{fig:distribution_probabilities}
\end{figure}

\begin{figure}
    \centering
    \includegraphics[width=0.4\textwidth]{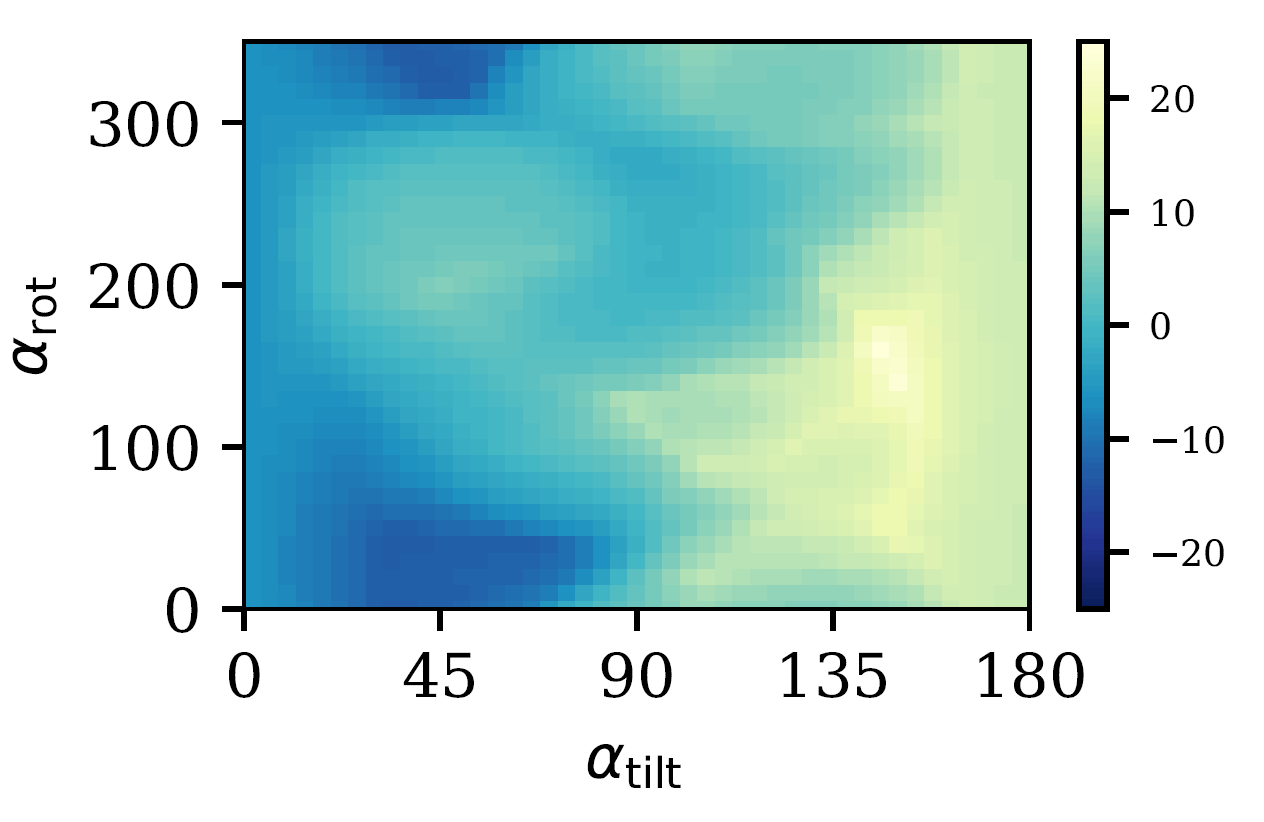}
    \includegraphics[width=0.4\textwidth]{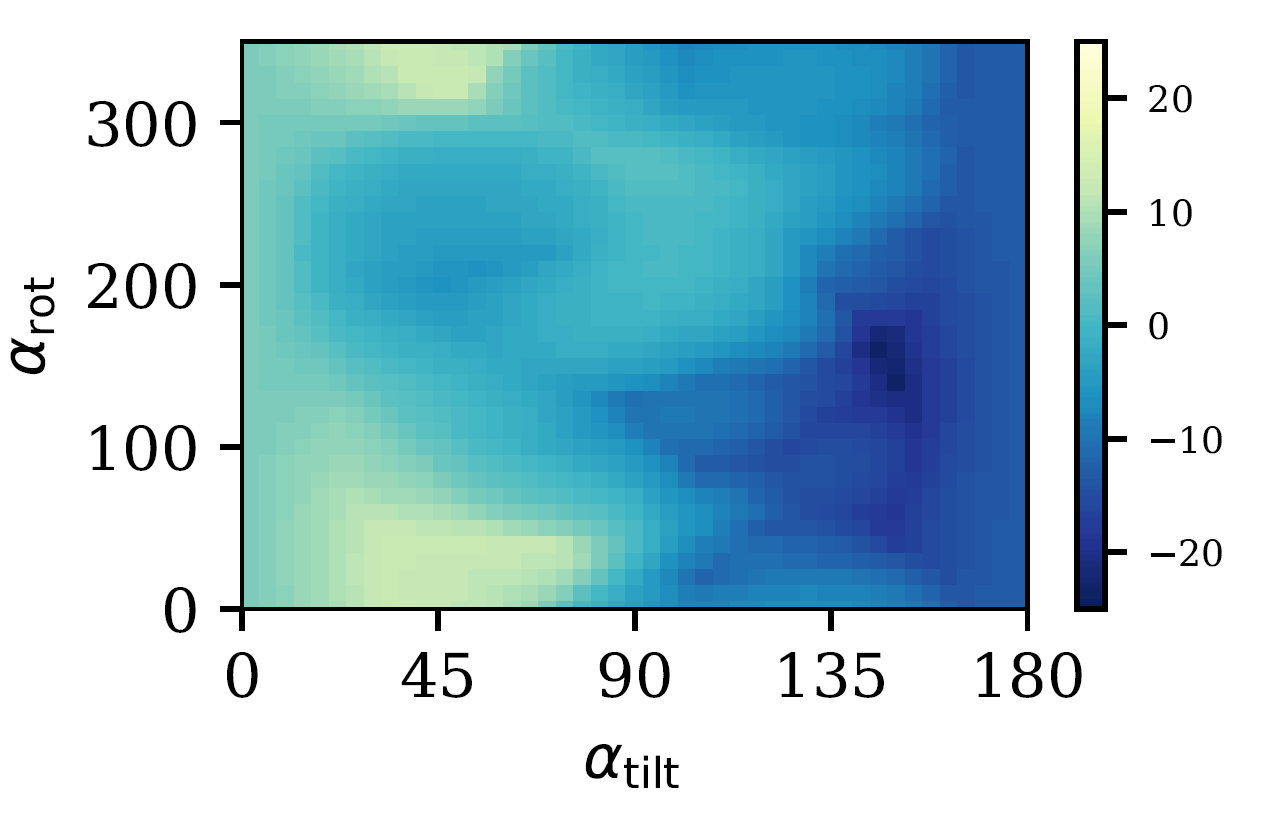}
    \caption{Interaction energy ($\Delta G_{int}$, in kcal/mol) at different orientations for $\phi_0$=-1 V (left) and $\phi_0$=1 V (right).}
    \label{fig:deltaGint_orientations}
\end{figure}

Table \ref{table:trypsin} contains the preferred orientation for different external potentials ($\phi_0$ = $\pm$1 and 0 V), with the corresponding interaction energy $\Delta G_{int}$. 
We can see that the interaction is attractive for the preferred orientation, however, it is weaker for a negative applied potential, and negligible when there is no field. This is further confirmed by Fig. \ref{fig:deltaGint_orientations}, that displays both attractive and repulsive interactions. Also, Fig. \ref{fig:distribution_probabilities} shows that the probability without an external field is completely random, whereas it localizes to specific orientations with the external potential turned on, with a much sharper distribution for $\phi_0$=1V. 

\begin{figure}
    \centering
    \includegraphics[width=0.85\textwidth]{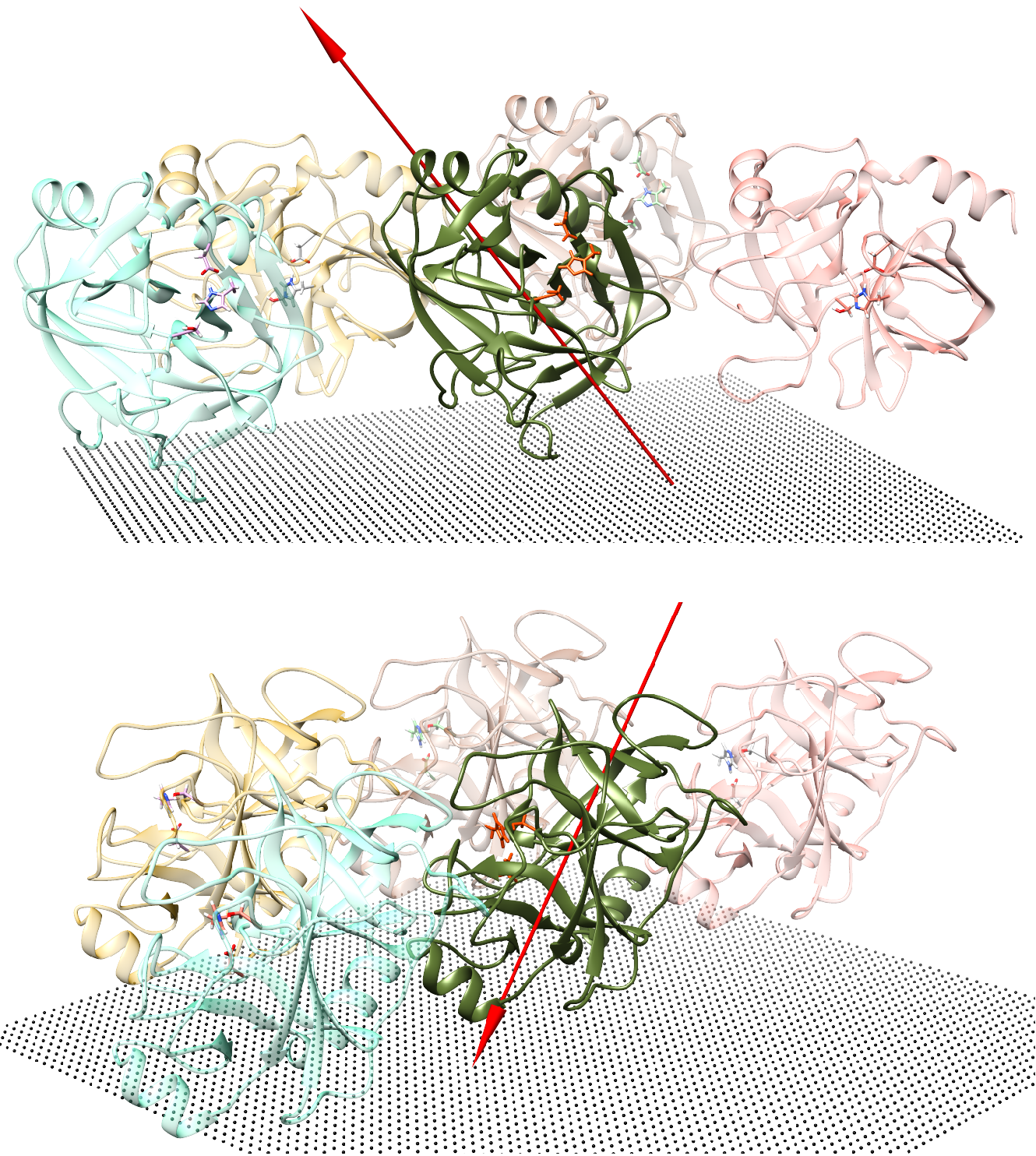}
    \caption{Preferred orientations for negative (top) and positive (bottom) applied potentials. The active site in each case is marked in red. Note that our simulations considered only one trypsin.}
    \label{fig:preferred_orientation}
\end{figure}

The active sites are located at HIS57, ASP102, and SER195. Table \ref{table:access_factor} shows their average accessibility factor ($\bar{f}$ in Eq. \eqref{eq:avg_access}), and Fig. \ref{fig:preferred_orientation} is a detailed view of trypsin in its preferred orientation for each case, with the active sites highlighted. 

We studied the effect of salt concentration on the trypsin-surface interaction, in particular, looking at the variations of energy and orientation. These results are summarized by Figure \ref{fig:Gint_kappa}, and further detailed in Table S4 of the Supplementary Material. 

\begin{table} 
\centering
\caption{PyGBe parameters for trypsin runs.}
\label{table:orientation_trypsin_param}
\begin{tabular}{ccccccc}
\hline
\multicolumn{3}{c}{\# Gauss points} & \multicolumn{3}{c}{Treecode} & GMRES     \\ \hline
$K$    & $K_{fine}$    & $N_{k}$    & $N_{crit}$  & P   & $\theta$ & tol.      \\ \hline
4      & 19            & 9          & 500         & 6  & 0.5      & $10^{-5}$ \\ \hline
\end{tabular}
\end{table}

\begin{table}
 \centering
  \caption{Preferred orientation of trypsin adsorbed on a charged surface under an electric field with $\kappa=0.175$ and $\sigma_0=0$.}
  \label{table:trypsin}
  \begin{tabular}{cccc}
    \hline
     $\phi_0$ & $\alpha_{tilt}$ & $\alpha_{rot}$ & $\Delta G_{int}$ \\
    V    &               &               & kcal/mol      \\
    \hline
    -1 & 36  & 340 & -12.85 \\
    0  & 96  & 200 & -0.049 \\
    1  & 144 & 160 & -23.65 \\
    \hline
  \end{tabular}
\end{table}

\begin{table}
 \centering
  \caption{Average accessibility factor ($\bar{f}$) of active sites, computed with Eq. \eqref{eq:avg_access}}
  \label{table:access_factor}
  \begin{tabular}{c|ccc|ccc|ccc}
        & \multicolumn{3}{c|}{HIS57} &  \multicolumn{3}{c}{ASP102} & \multicolumn{3}{|c}{SER195} \\
     $\phi_0$ (V) & -1 & 0 & 1 &  -1 & 0 & 1 &  -1 & 0 & 1  \\
    \hline
    $\bar{f}$ & 0.62 & 0.53 & 0.52 & 0.73 & 0.53 & 0.38 & 0.53 & 0.53 & 0.58 \\

  \end{tabular}
\end{table}

\begin{figure}
    \centering
    \includegraphics[width=0.45\textwidth]{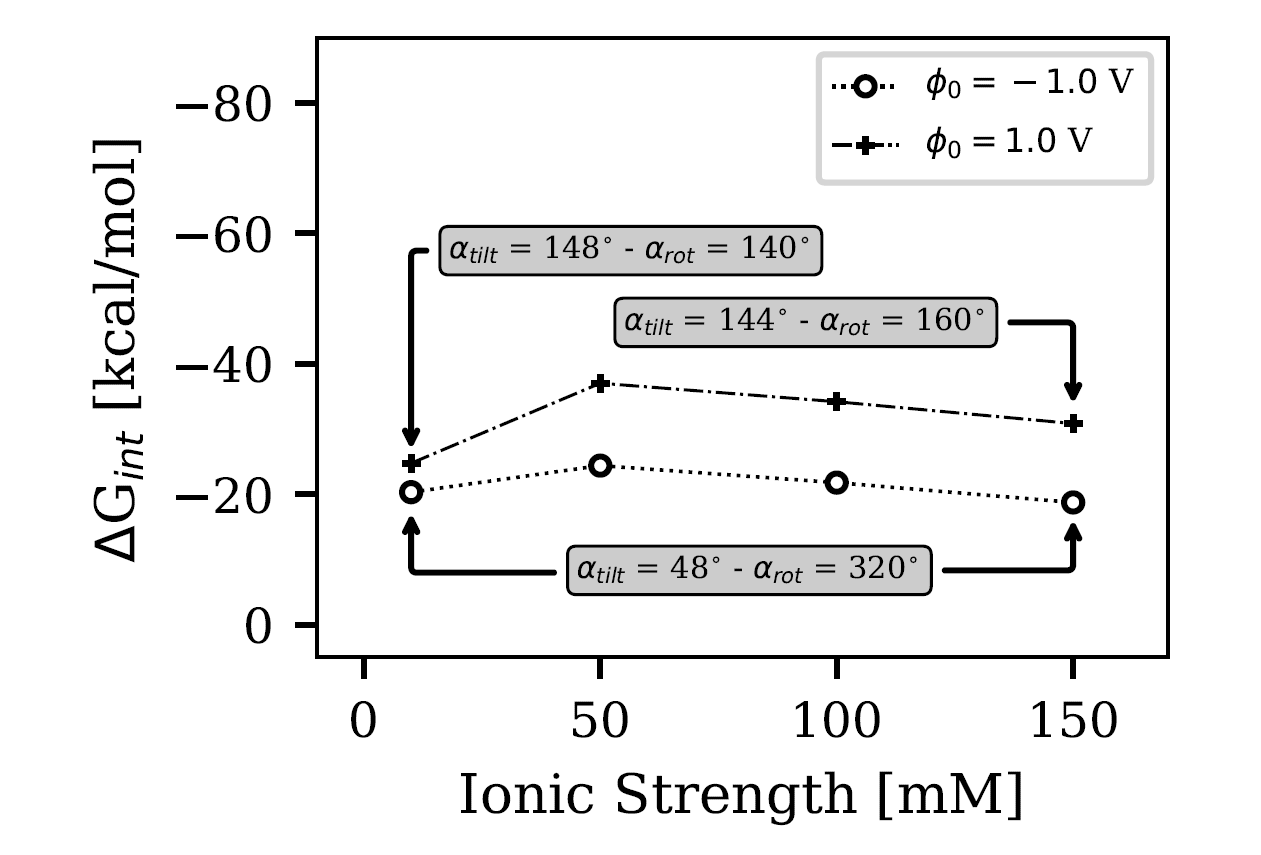}
    \caption{Interaction energies and their respective tilt/rot angles for trypsin adsorbed on an uncharged surface under an electric field, as a function of the ionic strength. Exact values are detailed in Table S4 of the Supplementary Material.}
    \label{fig:Gint_kappa}
\end{figure}

%% file: discussion.tex
\subsection{Validity of an electrostatic model to explain the experimental observations}

Emulating an experimental setup with computer simulations is a challenging task. The real behavior of proteins during experiments is usually far more complex than most models can describe. Specialized models developed to account for complete phenomena are generally too expensive for researchers to obtain useful simulations. While the present case is not an exception, we have identified different mechanisms that explain experimental observations, where simulations can be used to describe specific aspects of the behavior. 

 To begin with, in the experiments, the external field polarized the electrodes, inducing a  surface charge on them. However, we kept $\sigma_0$=0 C/m$^2$ throughout, due to the low density of functional groups present on the surface of the selected carbon electrodes. 

Also, the Poisson-Boltzmann model only accounts for electrostatic effects, which is just one component of the trypsin-electrode interaction. In fact, for hydrophobic surfaces like the carbon electrodes, this attractive interaction is typically dominated by the hydrophobic effect, through  dewetting of the electrode-protein interface~\cite{doi:10.1021/la062238s}. Moreover, the behavior reported on other carbon substrates \cite{HUANG2020122285}, the low density of functional groups of the substrates \cite{REED202090}, and the concentration of buffer used (0.1M phosphate, pH=8), also support the general notion that the adsorption process is mostly driven by hydrophobic interactions \cite{doi:10.1021/la062238s}. The latter is confirmed by the catalytic activity results without external potential from the middle bar of Fig. \ref{fig:figactivity}. In that case, the activity results indicate good coverage of trypsin on the surface. Then, the difference in catalytic activity for positive and negative external fields (two side bars in Fig. \ref{fig:figactivity}) can be attributed to the electrostatic contributions, as it is based on the response of trypsin to the external electric field, and how this affects the orientation at adsorption. This is where Poisson-Boltzmann simulations become particularly useful.

The Poisson-Boltzmann model by itself is not free of approximations. First, it considers a rigid molecular geometry. This may seem like a major limitation, specially considering carbon surfaces may induce conformational changes on trypsin \cite{HUANG2020122285}. However, these changes would happen in all conditions described in Fig.~\ref{fig:figactivity}, making them comparable. Moreover, it is important to consider that such changes are likely to span beyond the timescale selected for the experiments \cite{NORDE2000259, la990416u}, and that an external field may provide a stabilizing effect on secondary structures \cite{budi2005electric}. Also, the simulations were performed at a single protonation state, valid for an isolated trypsin at the corresponding pH, however, the surface may induce pKa shifts. Even though this effect, known as charge regulation \cite{ninham1971electrostatic,lund2013charge}, has an impact on electrostatics, its influence on the adsorption orientation can be neglected \cite{boubeta2018electrostatically}. Finally, the standard Poisson-Boltzmann equation does not take into account ion-specific effects, such as ion-ion correlation, but all experiments were performed at the same ionic conditions, allowing us to discard the impact of these effects in the observed catalytic activity difference. 

It is important to note that this model has limitations regarding the linear approximation of the Poisson-Boltzmann equation. The linearization yields acceptable results in most protein systems, even when the electrostatic potential is above the linearization condition \cite{fogolari1999biomolecular,fogolari2002poisson}, but it may become a problem for highly charged systems, like nucleic acids.

\subsection{The effect of an external field on the catalytic activity}

There are a number of possible causes to explain the differences in catalytic activity described by Fig. \ref{fig:figactivity}, which not only refer to the amount of enzymes adsorbed on the surface, but also their conformation \cite{doi:10.1021/ac802068n}. Considering the hydrophobic contributions to the adsorption process, and the enhancement provided by favorable electrostatic interaction free energies in Table \ref{table:trypsin} and Fig. \ref{fig:deltaGint_orientations}, we do not expect to see major differences in trypsin coverage on the surface of the electrodes prepared under different applied potentials. This is further supported by comparing $\Delta G_{int}$ with positive and negative electric fields in Fig. \ref{fig:deltaGint_orientations}. In that case, the surface has higher electrostatic affinity with trypsin when the electric field is positive (see Table \ref{table:trypsin}). This suggests that, if trypsin coverage plays a role in the overall catalytic activity, the latter would be higher with a positive electric field. However, Fig. \ref{fig:figactivity} shows the opposite result: a negative electric field has higher activity. These observations are aligned with experimental results of trypsin adsorbed on metallic surfaces \cite{htwe2018adsorption}, where the coverage is independent of the applied field for a pH below the isoelectric point of the molecule.

Other researchers have proposed that local changes of the pH, due to electrolysis of the solvent \cite{D0CC06190E}, could also induce aggregation of proteins and affect the adsorbed amount. However, the selected carbon electrodes showed rather poor electrochemical activity towards the reduction O$_{2}$. In addition, under these conditions the application of a negative potential (-1.0 V vs Ag/AgCl/KCl$_\text{SAT}$) would lead to a local \textit{increase} in the pH, bringing the protein closer to its isoelectric point (10.5) \cite{ZHANG2010781} and thus further decreasing the general influence of (attractive) electrostatic interactions. Moreover, the application of an external electric field can also induce the accumulation of proteins due to a polarization effect, but this is a rather slow process that would not render significant differences in our electrodes \cite{NORDE2000259,norde2007my}.

This motivates us to consider the effect of the external electric field on the orientation as the driving mechanism behind the catalytic activity difference in Fig. \ref{fig:figactivity}.

The position of the active sites for the preferred orientations with positive or negative external fields are different (see Fig. \ref{fig:preferred_orientation}), however, none of them is evidently more accessible. Moreover, comparing only the preferred orientations is misleading, as the probabilities with $\phi_0$=-1 V are much more spread out than the $\phi_0$=1 V case (notice the difference in the span of the colorbars in Fig. \ref{fig:distribution_probabilities}). The average accessibility ($\bar{f}$) in Table \ref{table:access_factor} gives a more complete notion of the effect of an external field on the catalytic activity. For a random orientation distribution, $\bar{f}\approx 0.5$, which is approximately what we obtain without an external field. When $\phi_0$=1 V, $\bar{f}$ decreases for 2 out of 3 sites, indicating that it is harder for an antigen to bind to trypsin. On the other hand, $\bar{f}$ increases for $\phi_0$=-1 V, which results in more available binding sites, and leads to an enhanced catalytic activity. This rationale is aligned with the experimental observations in Fig. \ref{fig:figactivity}.

The external field also has an effect on the shape of the orientation distribution. In particular, this distribution is narrower with the electric field is applied compared to the case without an external field (Fig. \ref{fig:distribution_probabilities}). This is the result of an energy landscape with a deeper well when the external field is present, and yields less random orientations at adsorption. On the other hand, with an external field, energies are more negative (see Table \ref{table:trypsin}), and hence, more favorable, regardless of the sign of the external field. Comparing the cases with positive and negative applied potential in Fig. \ref{fig:deltaGint_orientations}, we can see that the interaction is more favorable when $\phi_0$ is positive. This is also supported by the average interaction energy, computed as the sum of $\Delta G_{int}$ times the probability at each conformation, which is -12.28 kcal/mol with $\phi_0$=-1 V, and -23.26 kcal/mol for $\phi_0$=1 V.

Fig. \ref{fig:deltaGint_orientations} shows orientations with repulsive and attractive electrostatic interactions. Then, the electrode may repel trypsin if it approaches in certain orientations, or force the molecule to rotate and conform to one of the most likely configurations. Similar numerical techniques based on the boundary element method could be used to study the hydrodynamics in the trypsin rotation as it approaches the electrode \cite{aragon2011recent}, however, we leave such study for future work.

\subsection{Effect of ionic strength}

Figure \ref{fig:Gint_kappa} presents the interaction free energy for the preferred orientation, as a function of ionic strength. The angles with highest probability change only slightly, but the impact on the interaction energy is more substantial. With either a negative or positive electric field, the electrostatic contribution initially becomes more favorable with increasing ionic strength, which may seem counter intuitive because of the increased screening. Then, however, the ionic shielding effect becomes more important and decreases the electrostatic interaction, which agrees with similar studies \cite{mulheran2016steering}. The results also show that there is a stronger attractive interaction with a positive electric field compared to a negative electric field, and this difference is larger at higher salt concentrations. Considering that trypsin is positively charged, this is an indication that local interactions between the surface and the protein are more important than the monopole-type charge interaction. If the net charge were dominating, the positively charged trypsin would be pushed away from the surface by a positive electric field.

%% file: conclusions.tex
Being able to control the orientation of proteins adsorbed on surfaces is key in a variety of applications, such as biosensors and biocatalysis. In this work, we provide evidence that applying an electric field as a protein adsorbs on a surface has an important effect on the orientation and the subsequent catalytic activity. Even though the adsorption process is dominated by hydrophobic interactions, we were able to identify that electrostatic interactions can affect the catalytic activity difference via the promotion of certain orientations on the surface. In particular, we argue that other effects that are usually critical to consider, such as conformational changes, charge regulation, and trypsin coverage on the surface, can play a secondary role for the proposed analysis. Towards a rational use of electrostatics to promote the orientation of certain proteins, we present an efficient computational model based on the Poisson-Boltzmann equation. The model not only enables studying the electrostatics of this system but also predict the most likely orientations, as a function of the external potential. Our numerical experiments support the experimental observations, as the active sites are, on average, more exposed to the solvent with a negative applied potential, as opposed to when the field was positive. This motivates the use of computational studies with implicit-solvent models and electrostatics in the application of electric fields for fine tuning of molecular immobilization on surfaces. As future work, we plan to use this approach to the aid the design of more sensitive surfaces in biotechnological settings, and study crowding effects by including more than one molecule.
